\documentclass[journal,romanappendices]{IEEEtran}

\usepackage{graphicx,amsfonts,amsmath,cite,comment,enumitem,hyperref,mathtools,algorithm,algpseudocode,setspace,amsthm,subcaption}

\usepackage{tikz,pgfplots}
\usetikzlibrary{arrows, arrows.meta, backgrounds, calc, intersections, decorations.markings, decorations.pathreplacing, positioning, shapes}
\pgfplotsset{compat=1.17}

\usepackage{caption}
\usepackage{titlesec}
\let\subparagraph\relax
\titlespacing{\section}{0pt}{8pt plus 2pt minus 1pt}{4pt plus 1pt minus 1pt} 
\titlespacing{\subsection}{0pt}{6pt plus 2pt minus 1pt}{2pt plus 1pt minus 1pt} 
\newcommand{\Trans}{{\mathrm{T}}}
\newcommand{\Herm}{{\mathrm{H}}}
\newcommand{\trace}{{\mathrm{tr}}}
\newcommand{\VEC}{\mathrm{vec}}
\newtheorem{proposition}{Proposition}

\renewcommand{\a}{\mathbf{a}}

\newcommand{\e}{\mathbf{e}}

\newcommand{\g}{\mathbf{g}}
\newcommand{\h}{\mathbf{h}}

\newcommand{\m}{\mathbf{m}}

\newcommand{\p}{\mathbf{p}}

\renewcommand{\r}{\mathbf{r}}
\newcommand{\s}{\mathbf{s}}

\renewcommand{\u}{\mathbf{u}}

\newcommand{\w}{\mathbf{w}}
\newcommand{\x}{\mathbf{x}}
\newcommand{\y}{\mathbf{y}}
\newcommand{\z}{\mathbf{z}}

\newcommand{\0}{\mathbf{0}}

\newcommand{\A}{\mathbf{A}}
\newcommand{\B}{\mathbf{B}}
\newcommand{\C}{\mathbf{C}}
\newcommand{\D}{\mathbf{D}}

\newcommand{\F}{\mathbf{F}}

\renewcommand{\H}{\mathbf{H}}
\newcommand{\I}{\mathbf{I}}
\newcommand{\J}{\mathbf{J}}

\renewcommand{\P}{\mathbf{P}}
\newcommand{\Q}{\mathbf{Q}}
\newcommand{\R}{\mathbf{R}}

\newcommand{\U}{\mathbf{U}}
\newcommand{\V}{\mathbf{V}}
\newcommand{\W}{\mathbf{W}}
\newcommand{\X}{\mathbf{X}}

\newcommand{\Z}{\mathbf{Z}}

\newcommand{\varepsilonb}{\boldsymbol{\varepsilon}}

\newcommand{\etab}{\boldsymbol{\eta}}

\newcommand{\mub}{\boldsymbol{\mu}}

\newcommand{\omegab}{\boldsymbol{\omega}}

\newcommand{\Sigmab}{\mathbf{\Sigma}}

\newcommand{\setC}{\mathcal{C}}

\newcommand{\setN}{\mathcal{N}}

\newcommand{\Compl}{\mbox{$\mathbb{C}$}}
\newcommand{\Real}{\mbox{$\mathbb{R}$}}

\newcommand{\argmin}{\operatornamewithlimits{argmin}}
\newcommand{\blkdiag}{\mathrm{blkdiag}}

\newcommand{\Diag}{\mathrm{Diag}}

\newcommand{\Exp}{\mathbb{E}}
\newcommand{\herm}{\mathrm{H}}
\renewcommand{\Im}{\mathrm{Im}}

\renewcommand{\Re}{\mathrm{Re}}
\newcommand{\sgn}{\mathrm{sgn}}

\newtheorem{remark}{Remark}

\definecolor{oulu_blue}{HTML}{23408F}
\definecolor{oulu_green}{HTML}{39B54A}

\definecolor{red}{rgb}{1,0,0}
\definecolor{red_magenta}{rgb}{1,0,0.5}
\definecolor{magenta}{rgb}{1,0,1}
\definecolor{blue_magenta}{rgb}{0.5,0,1}
\definecolor{blue}{rgb}{0,0,1}
\definecolor{blue_cyan}{rgb}{0,0.5,1}
\definecolor{cyan}{rgb}{0,1,1}
\definecolor{green_cyan}{rgb}{0,1,0.5}
\definecolor{green}{rgb}{0,1,0}
\definecolor{green_yellow}{rgb}{0.5,1,0}
\definecolor{yellow}{rgb}{1,1,0}
\definecolor{red_yellow}{rgb}{1,0.5,0}

\title{Nonlinear Sparse Bayesian Learning Methods with Application to Massive MIMO Channel Estimation with Hardware Impairments}
\author{Arttu~Arjas and Italo~Atzeni
\thanks{The authors are with the Centre for Wireless Communications, University of Oulu, Finland (e-mail: \{arttu.arjas, italo.atzeni\}@oulu.fi). Part of this work was presented at IEEE ICASSP 2025 \cite{arjas2025enhanced}. \\ \indent This work was supported by the Research Council of Finland (336449 Profi6, 348396 HIGH-6G, and 369116 6G~Flagship).}}

\begin{document}

\maketitle

\begin{abstract}
Accurate channel estimation is critical for realizing the performance gains of massive multiple-input multiple-output (MIMO) systems. Traditional approaches to channel estimation typically assume ideal receiver hardware and linear signal models. However, practical receivers suffer from impairments such as nonlinearities in the low-noise amplifiers and quantization errors, which invalidate standard model assumptions and degrade the estimation accuracy. In this work, we propose a nonlinear channel estimation framework that models the distortion function arising from hardware impairments using Gaussian process (GP) regression while leveraging the inherent sparsity of massive MIMO channels. First, we form a GP-based surrogate of the distortion function, employing pseudo-inputs to reduce the computational complexity. Then, we integrate the GP-based surrogate of the distortion function into newly developed enhanced sparse Bayesian learning (SBL) methods, enabling distortion-aware sparse channel estimation. Specifically, we propose two nonlinear SBL methods based on distinct optimization objectives, each offering a different trade-off between estimation accuracy and computational complexity. Numerical results demonstrate significant gains over the Bussgang linear minimum mean squared error estimator and linear SBL, particularly under strong distortion and at high signal-to-noise ratio.
\end{abstract}


\section{Introduction} \label{sec:intro}

Channel estimation is crucial to enable beamforming design in multiple-input multiple-output (MIMO) systems. It becomes even more important for massive MIMO, where large antenna arrays allow for transmit/receive beamforming with extreme spatial resolutions \cite{Bjo17,Raj20}. Channel estimation is usually carried out via uplink pilots, where the user equipments (UEs) send predetermined pilot signals to the base station (BS) to probe the channels. When the channels are Gaussian, the linear minimum mean squared error (LMMSE) channel estimator can be shown to be optimal among the linear estimators. To reduce pilot overhead, compressed sensing techniques such as sparse Bayesian learning (SBL), matching pursuit, or $\ell_1$-norm regularization can be used to estimate the channels \cite{berger2010application,lee2016channel,cotter2002sparse,prasad2014joint,pedersen2012application,schniter2014channel}. These methods exploit the angular sparsity of the channels arising, for instance, when there is a limited number of scatterers between the transmitter and the receiver, which causes most of the received signal to come from few channel~paths.

The aforementioned approaches to channel estimation typically assume ideal hardware at the receiver, which is unrealistic in real-world systems. In practice, the receiver hardware is non-ideal, creating impairments such as nonlinearities in the low-noise amplifiers (LNAs), I/Q imbalance, phase noise, and quantization errors \cite{bjornson2014massive}. Although these impairments can be partly mitigated by compensation algorithms \cite{schenk2008rf}, ignoring the residual impairments in the channel estimation leads to decreased estimation accuracy \cite{bjornson2014massive}. In practical systems, channel reciprocity is disrupted due to non-ideal transceiver hardware, which introduces different impairments at the transmitter and receiver. These hardware asymmetries lead to uplink and downlink channels that are no longer simple transposes of one another \cite{mokhtari2019survey}. To overcome this, one must separately estimate the propagation channel, which excludes the transceiver hardware, and model the hardware impairments separately. The resulting nonlinear mapping can then be accounted for in the data detection stage. Despite their practical relevance, hardware impairments are rarely modeled explicitly in channel estimation frameworks, creating a gap between theory and practice that motivates the present work.

In signal processing, a popular approach to analyze nonlinearities is to use the Bussgang decomposition \cite{bussgang1952}, which allows to express the output of a nonlinear function as a linearly scaled version of the input plus an uncorrelated distortion term. This statistical tool can be utilized to analyze communication systems under hardware impairments \cite{demir2020bussgang} and design distortion-aware extensions of common linear channel estimation and data detection methods (such as LMMSE) based on the first- and second-order statistics of the distortion term \cite{li2017channel,bjornson2018hardware,Din25}. However, these methods assume perfect covariance and cross-covariance information of the input and output. Moreover, approaches that directly model the nonlinear transformation itself can enable more accurate compensation under significant hardware impairments. This motivates the need for a unified framework for nonlinear estimation and analysis in wireless systems, as recently highlighted in \cite{chafii2023twelve}.

In this work, we propose to model the hardware impairments using Gaussian processes (GPs), a data-driven tool that can be utilized to model nonlinear functions; we refer to \cite{williams2006gaussian} for a thorough introduction to GPs. In general, GPs can be used to learn nonlinear functions from paired input-output training data. Given the data, the GP can be evaluated at any input point not included in the training data using Bayes' formula. The properties of a GP (e.g., smoothness and scale of the nonlinearity) are governed by a covariance function, also referred to as kernel, which defines the similarity between two inputs, usually based on their mutual distance. Compared with parametric models, GPs are more flexible as they do not assume a fixed functional form but rather assign a prior distribution over functions. By incorporating prior assumptions on the function to be learned, GPs have been shown to perform well even with limited data \cite{snoek2012practical}. In contrast, methods based on deep learning typically lack interpretability and require large training datasets due to the substantial number of parameters involved \cite{soltani2019deep,sun2017revisiting,gao2023deep}. Nonetheless, they offer strong representational flexibility and can learn highly complex, data-driven nonlinear relationships once adequately trained, making them powerful tools when sufficient data is available. Alternatively, support vector machines can learn nonlinear relations by projecting the input data onto a high-dimensional space using kernels, and have been applied to channel estimation in \cite{sanchez2004svm,Ngu21}. However, with nonlinear kernels, the recovery of the channels becomes difficult since the estimate lies in the reproducing kernel Hilbert space and cannot be directly mapped back to the original channel space in closed form. GPs typically require tuning only few hyperparameters, some of which can be automatically selected using approaches such as maximum likelihood (ML). A well-known limitation of standard GPs is their cubic computational cost with respect to the number of training samples \cite{williams2006gaussian}. To address this, several techniques for reducing the computational complexity have been proposed, including low-rank matrix approximations \cite{williams2000using, ambikasaran2015fast}. A promising approach to enhance the scalability of GPs is using pseudo-inputs \cite{snelson2005sparse}, which summarize the training data with a smaller set of representative points, thereby reducing the computational complexity.

\textbf{\textit{Contribution.}} Despite the significance of hardware impairments in real-world communication systems, existing channel estimation techniques predominantly assume ideal hardware. Neglecting hardware impairments not only leads to performance degradation but also fails to account for the resulting disruption of channel reciprocity. To bridge this gap, we propose a nonlinear channel estimation framework that models the distortion function arising from hardware impairments while leveraging the inherent sparsity of massive MIMO channels. The distortion function is replaced by a GP-based surrogate function learned from data, with pseudo-inputs employed to reduce the computational complexity. The GP-based surrogate function is then integrated into newly developed enhanced SBL methods, enabling distortion-aware sparse channel estimation. The main contributions are summarized as follows.
\begin{itemize}
    \item We consider a MIMO system characterized by channel sparsity and hardware impairments, with the LNA distortion serving as our main motivating example. We model the nonlinear distortion function using the GP regression framework to form a learned surrogate function that replaces the distortion function in the computations. A key advantage of this approach is that an explicit mathematical form of the distortion function is not required. Furthermore, we utilize pseudo-inputs that notably decrease the computational complexity related to the evaluation and differentiation of the GP-based surrogate function.
    \item We integrate the GP-based surrogate function into the SBL framework for sparse channel estimation. In this context, we develop enhanced SBL methods by introducing a vector of scales alongside the traditional weight vector. In the linear case, SBL iteratively solves a system of linear equations and subsequently updates the weights and scales. When the surrogate function is nonlinear, this extends to (approximately) solving a system of nonlinear equations before performing the same updates. Specifically, we propose two enhanced SBL methods that incorporate the GP-based surrogate function: one that maximizes the marginal posterior density over the weights and scales, and another that maximizes the joint posterior density over the channel, weights, and scales. The proposed methods offer different trade-offs between estimation accuracy and computational complexity.
    \item We analyze the computational complexity of the proposed nonlinear estimation framework and investigate how the number of pseudo-inputs affects the estimation accuracy. Based on this, we provide practical guidelines for selecting the number and locations of the pseudo-inputs. In addition, we discuss implementation aspects such as step size adaptation and initialization.
    \item We extend the proposed nonlinear estimation framework to cover important special cases such as hybrid analog-digital beamforming and 1-bit analog-to-digital converters (ADCs). The former requires adapting only the surrogate function, while the latter entails modifying both the optimization objective and the surrogate function.
    \item Considering the LNA distortion as our main motivating example, we numerically investigate the performance of the proposed nonlinear estimation methods against parameters such as the signal-to-noise ratio (SNR), pilot length, number of antennas at the BS, number of channel paths, and strength of the LNA distortion. Our results show that the proposed nonlinear estimation framework significantly outperforms conventional methods such as least squares (LS), Bussgang LMMSE (BLMMSE), and linear SBL in terms of normalized mean squared error (NMSE) of the channel estimation, particularly under strong LNA distortion and at high SNR.
\end{itemize}

Part of this work was presented in our conference paper \cite{arjas2025enhanced}, which proposed enhanced SBL methods for channel estimation assuming ideal (i.e., linear and distortion-free) receiver hardware.

\textbf{\textit{Outline.}} The rest of the paper is organized as follows. Section~\ref{sec:sys_model} introduces the system model with hardware impairments. Sections~\ref{sec:gp} and~\ref{sec:estimation} present the proposed nonlinear estimation framework: first, Section~\ref{sec:gp} models the hardware impairments using GPs; then, Section~\ref{sec:estimation} develops two sparse channel estimation methods that are embedded into the GP-based framework. Section~\ref{sec:implementation} discusses the computational complexity and implementation aspects, whereas Section~\ref{sec:special_cases} presents extensions to hybrid analog-digital beamforming and 1-bit ADCs. Finally, section~\ref{sec:num} presents extensive numerical results, and Section~\ref{sec:concl} concludes the paper.

\textbf{\textit{Notation.}} Transpose, Hermitian transpose, and complex conjugate are denoted by $(\cdot)^\Trans$, $(\cdot)^\Herm$, and $(\cdot)^*$, respectively. Vectors and matrices are expressed by bold lowercase and uppercase letters, respectively. The $j$th element of a vector $\x$ without subscripts is denoted by $x_j$; if the vector has a subscript as $\x_\textrm{s}$, the element is expressed by $[\x_\textrm{s}]_j$. The $(j,l)$th element a matrix $\X$ without subscripts is denoted by $X_{jl}$; if the matrix has a subscript as $\X_\textrm{s}$, the element is expressed by $[\X_\textrm{s}]_{jl}$. The $j$th row of a matrix $\X$ is denoted by $\X_{j:}$, and the elements $j$ through $l$ of a vector $\x$ by $\x_{j:l}$. Diagonal and block-diagonal matrices are defined using $\Diag (\cdot)$ and $\blkdiag (\cdot)$, respectively. The elementwise (Hadamard) and Kronecker products are expressed by $\odot$ and $\otimes$, respectively. The sign function is denoted by $\sgn(\cdot)$. Proportionality is indicated by $\propto$. The circularly symmetric complex Gaussian distribution with mean $\m$ and covariance matrix $\C$ is denoted by $\mathcal{CN}(\m, \C)$, whereas the inverse-gamma distribution with shape $\gamma$ and scale $\beta$ is denoted by $\mathcal{IG}(\gamma, \beta)$. The probability density function of a random variable $x$ given another random variable $y$ is denoted by $p(x|y)$. Lastly, the imaginary unit is indicated by~$i$.

\section{System Model} \label{sec:sys_model}

In this section, we first describe the considered MIMO system model with hardware impairments. Then, we present the BLMMSE channel estimator, which will be used as a baseline in Section~\ref{sec:num}. Lastly, we introduce the proposed nonlinear estimation framework that will be developed in the following sections.

\subsection{Channel Model and Hardware Impairments} \label{sec:channelmodel}

We consider the problem of uplink channel estimation in a MIMO system where a BS with $M$ antennas serves $K$ single-antenna UEs. The UEs simultaneously transmit known pilot sequences of length $N$, which are collected in the pilot matrix $\P \in \Compl^{K \times N}$. The signal at the BS's antennas is given by\footnote{Under the narrowband (flat-fading) assumption, small residual timing and frequency offsets do not alter the channel representation, and standard timing-advance and frequency-tracking procedures ensure that the system operates within this regime.}
\begin{equation} \label{eq:MIMO}
    \Z = \sqrt{p} \H\P \in \Compl^{M \times N},
\end{equation}
where $p > 0$ is the transmit power, $\H = [\h_{1}, \ldots, \h_{K}] \in \Compl^{M \times K}$ is the channel matrix, and $\h_{k} \in \Compl^{M}$ denotes the channel of UE~$k$. We consider correlated Rayleigh fading such that $\h_k \sim \mathcal{CN}(\0, \C_{\h_k})$, $\forall k = 1, \dots, K$. Furthermore, we assume that the channels are sparse in some known domain and express the channel matrix as $\H = \F\U$, where $\F \in \Compl^{M \times M}$ is a transformation matrix and $\U \in \Compl^{M \times K}$ is a matrix with sparse columns representing the channels in the transformed domain. In this paper, we consider far-field propagation and a uniform linear array (ULA) at the BS: this gives rise to angular sparsity, so $\F$ is defined as the discrete Fourier transform (DFT) matrix. However, the proposed method and analysis are readily applicable to any transformation matrix. For convenience, we vectorize \eqref{eq:MIMO} as
\begin{equation} \label{eq:linearmodel}
    \z = \VEC(\sqrt{p} \H\P) = \VEC(\sqrt{p} \F\U\P) = \sqrt{p} \A\u \in \Compl^{MN},
\end{equation}
with $\A = \P^\Trans \otimes \F \in \Compl^{MN \times MK}$ and $\u = \VEC(\U) \in \Compl^{MK}$.

Assuming non-ideal receiver hardware at the BS, the latter observes the distorted and noisy signal
\begin{equation} \label{eq:systemmodel}
    \y = g(\z) + \e \in \Compl^{MN},
\end{equation}
where $g(\z) = [g_1(z_1), \dots, g_{MN}(z_{MN})]^\Trans \in \Compl^{M N}$, with $g_j: \Compl \to \Compl$, $\forall j = 1, \dots, MN$, is a nonlinear function that models the hardware impairments \cite{bjornson2018hardware} and $\e \sim~\mathcal{CN}(\0, \sigma^2 \I_{MN})$ is a vector of additive white Gaussian noise (AWGN). Our methodology restricts $g$ to be smooth, which does not allow to capture discontinuous distortions such as quantization effects. This limitation can be addressed by adapting the likelihood function: in Section~\ref{sec:special_cases_1bit}, we follow this approach to handle the quantization distortion introduced by 1-bit ADCs.

\vspace{-2mm}

\begin{remark}
{\rm In practice, noise is introduced at the receiver both before and after the distortion function $g$. To justify the additive Gaussian model in \eqref{eq:systemmodel}, let $\e_1 \in \Compl^{MN}$ and $\e_2 \in \Compl^{MN}$ denote the pre- and post-nonlinearity AWGN terms, respectively. Using a first-order expansion, the effect of $\e_1$ passing through the distortion function can be approximated as $g(\A\u+\e_1)\approx g(\A\u)+\J_g(\A\u)\e_1$, where $\J_g(\A\u) \in \Compl^{MN \times MN}$ is diagonal since the nonlinearity acts elementwise. This yields an effective noise term that is approximately Gaussian for the mild nonlinearities of interest, and combining it with $\e_2$ leads to \eqref{eq:systemmodel}. This model captures the dominant distortion effects while keeping the inference analytically tractable and computationally efficient.}
\end{remark}

\vspace{-1mm}

Although $g(\cdot)$ might be partially known a priori, it is beneficial to learn it from data due to the somewhat unpredictable behavior of some hardware components.  Then, the estimated distortion function can be used to compensate for the hardware impairments at the receiver. To this end, \textit{we propose a nonlinear estimation framework that jointly estimates the distortion function $g (\cdot)$ and the sparse channel $\u$}.

Hardware impairments at the receiver originate, for example, from the LNAs and low-resolution ADCs. In \cite{bjornson2018hardware,schenk2008rf}, the LNA distortion is modeled by a differentiable (in the sense of Wirtinger derivatives) third-order nonlinear function, i.e.,
\begin{equation} \label{eq:lnadistortion}
    g_j(z_j) = z_j - a_j|z_j|^2 z_j, ~ \forall j = 1, \dots, MN,
\end{equation}
where $a_j > 0$ depends on the circuit technology and on the normalization of the LNA's output power. This is modeled in \cite{bjornson2018hardware,ericsson2016} as
\begin{equation}
    a_j = \frac{\alpha}{b_\textrm{off}\mathbb{E}[|z_j|^2]},
\end{equation}
where $\alpha > 0$ dictates the strength of the nonlinearity and $b_\textrm{off} \geq 1$ is a parameter chosen to limit the risk of clipping. On the other hand, the ADC distortion is characterized by a discontinuous quantization function. In this work, we consider the LNA distortion as our main motivating example, although we also extend our methodology to handle additional quantization distortion from 1-bit ADCs (see Section~\ref{sec:special_cases_1bit}).

\subsection{BLMMSE Channel Estimator} \label{sec:BLMMSE}

The Bussgang decomposition is a statistical tool that allows to reformulate a nonlinear function as a linear function with identical first- and second-order statistics. It can be used to analyze nonlinearities caused by hardware impairments \cite{bussgang1952,demir2020bussgang} as well as to design distortion-aware channel estimation and data detection methods \cite{li2017channel,Ngu21a,Din25}. In Section~\ref{sec:num}, we will use the BLMMSE channel estimator as a baseline for our proposed method. The Bussgang decomposition builds upon the Bussgang theorem to express a nonlinearly distorted signal as a linear function of the input summed with a distortion term that is uncorrelated with the input.

Applying the Bussgang decomposition to \eqref{eq:systemmodel} yields
\begin{equation}
    \y = \D \z + \etab + \e,
\end{equation}
where $\D = \mathbb{E}[g(\z)\z^\Herm]\mathbb{E}[\z\z^\Herm]^{-1} \in \Compl^{MN \times MN}$ is the Bussgang gain and $\etab = g(\z) - \D \z \in \Compl^{MN}$ is the zero-mean, non-Gaussian distortion term with covariance matrix $\C_{\etab} = \Exp[\etab \etab^{\herm}] \in \Compl^{MN \times MN}$. Since we assume $\h_k \sim \mathcal{CN}(\0, \C_{\h_k})$, $\forall k = 1, \dots, K$, the BLMMSE estimator for the vectorized angular channel is given by
\begin{equation} \label{eq:BLMMSE}
    \u_{\textrm{BLMMSE}} = \C_{\u} \A^\Herm \D \C_{\y}^{-1} \y \in \Compl^{MK},
\end{equation}
with
\begin{align}
\C_{\y} & = \Exp[\y \y^{\Herm}] \nonumber \\
& = \D \A \C_{\u} \A^\Herm \D^\Herm + \C_{\etab} + \sigma^2 \I_{MN} \in \Compl^{MN \times MN}, \\
\C_{\u} & = \Exp[\u \u^{\Herm}] \nonumber \\
& = \blkdiag(\F^\Herm \C_{\h_1} \F, \dots, \F^\Herm \C_{\h_K} \F) \in \Compl^{MK \times MK}.
\end{align}
The Bussgang gain $\mathbf{D}$ is diagonal because $g$ acts elementwise and the input $\mathbf{z}$ is Gaussian \cite{demir2020bussgang}. By Stein's lemma, the cross-correlation satisfies $\mathbb{E}[g(\mathbf{z})\mathbf{z}^\Herm] = \mathbf{D}\,\mathbf{C}_{\mathbf{z}}$, which implies $\mathbf{D} = \mathbb{E}[g(\mathbf{z})\mathbf{z}^\Herm]\mathbf{C}_{\mathbf{z}}^{-1}$ being diagonal. When the distortion function is given by \eqref{eq:lnadistortion}, the diagonal elements of $\D$ are given by
\begin{equation}
    D_{jj} = 1 - 2a_j[\C_{\z}]_{jj}, ~ \forall j = 1, \dots, MN,
\end{equation}
with $\C_{\z} = \Exp[\z \z^{\Herm}] = \A \C_{\u} \A^\Herm \in \Compl^{MN \times MN}$. Moreover, $\C_{\etab}$ has the form \cite{bjornson2018hardware}
\begin{equation}
    \C_{\etab} = 2\R(\C_{\z} \odot \C_{\z}^* \odot \C_{\z})\R,
\end{equation}
with $\R = \Diag(a_1, \dots, a_{MN}) \in \Compl^{MN \times MN}$. All the quantities required to compute the BLMMSE estimate in \eqref{eq:BLMMSE} follow from the channel covariance matrices $\C_{\h_k},$ $\forall k = 1, \dots, K$, and $\D$ additionally depends on the distortion parameters $a_j$, $\forall j = 1, \dots, MN$. In practice, the channel covariance matrices can be estimated from pilots and the distortion parameters from calibration measurements. In our numerical results, we assume these quantities to be known.

\subsection{Proposed Nonlinear Estimation Framework}

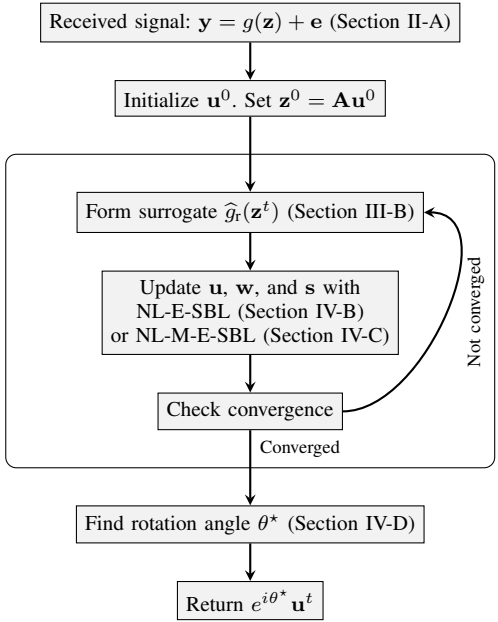
\begin{figure}[!t]
    \centering


\tikzstyle{process} = [rectangle, draw, fill=gray!10, align=center]
\tikzstyle{arrow} = [thick,->,>=stealth]

\centering
\begin{tikzpicture}[every node/.style={font=\footnotesize}]

\node (receive) [process] {Received signal: $\y = g(\z) + \e$ (Section~\ref{sec:channelmodel})};
\node (initialize) [process, below=0.5cm of receive] {Initialize $\u^0$. Set $\z^0 = \A \u^0$};
\node (surrogate) [process, below=1cm of initialize] {Form surrogate $\widehat{g}_\textrm{r}(\z^t)$ (Section~\ref{sec:gp_dimred})};
\node (update) [process, below=0.5cm of surrogate] {Update $\u$, $\w$, and $\s$ with \\ NL-E-SBL (Section~\ref{sec:NL-E-SBL}) \\ or NL-M-E-SBL (Section~\ref{sec:NL-M-E-SBL})};
\node (convergence) [process, below=0.5cm of update] {Check convergence};
\node (postprocess) [process, below=1cm of convergence] {Find rotation angle $\theta^\star$ (Section~\ref{sec:rotation})};
\node (return) [process, below=0.5cm of postprocess] {Return $e^{i\theta^\star}\u^t$};
 
\draw [arrow] (receive) -- (initialize);
\draw [arrow] (initialize) -- (surrogate);
\draw [arrow] (surrogate) -- (update);
\draw [arrow] (update) -- (convergence);
\draw [arrow] (convergence.east) to[out=0,in=0,looseness=1] node[midway, below, rotate=90, yshift=-1.5mm] {\scriptsize{Not converged}} (surrogate.east);
\draw [arrow] (convergence) --  node[midway, right, yshift=0.25cm] {\scriptsize{Converged}} (postprocess);
\draw [arrow] (postprocess) --  (return);

\draw[rounded corners] let \p1=($(surrogate.north)+(0cm,-0.25cm)$), \p2=($(convergence.south)+(-3.25cm,-0.5cm)$), \p3=($(convergence.south)+(3.25cm,-0.5cm)$), \p4=($(surrogate.north)+(0cm,0.5cm)$) in (\x2, \y1) -- (\x2, \y2) -- (\x3, \y3) -- (\x3, \y4) -- (\x2, \y4) -- (\x2, \y1);

\end{tikzpicture}

    \caption{Workflow diagram of the proposed nonlinear estimation framework.}
    \label{fig:diagram}
\end{figure}

To estimate the sparse channels while accounting for hardware impairments, we propose a nonlinear estimation framework that combines GPs with enhanced SBL methods. The core idea is to replace the unknown distortion function $g (\cdot)$ with a GP-based surrogate function $\widehat{g} (\cdot)$, which can be efficiently evaluated and differentiated for any input. A major advantage of this approach is that it eliminates the need for an explicit mathematical form of the distortion function, as $\widehat{g} (\cdot)$ is learned from data along with the channel estimation. We then develop enhanced SBL methods and embed them into the GP-based framework. As parameter estimation in SBL is accomplished by solving an optimization problem, integrating SBL with a nonlinear system model requires utilizing nonlinear optimization techniques. The proposed nonlinear estimation framework is summarized as a workflow diagram in Fig.~\ref{fig:diagram}, whereas its pseudocode is provided in Algorithm~\ref{alg:NL-SBL} (see Section~\ref{sec:estimation}). Next, Section~\ref{sec:gp} introduces GPs and describes the construction of the surrogate function, while Section~\ref{sec:estimation} presents two enhanced SBL methods for sparse channel estimation.

\section{Modeling Hardware Impairments Using GPs} \label{sec:gp}

In this work, we propose to use GPs to model the distortion function $g (\cdot)$. GPs are infinite-dimensional generalizations of Gaussian distributions, defined by a mean function $\mu: \Compl \to \Compl$ and a kernel $K: \Compl \times \Compl \to \Real$. The mean function models the average behavior of the GP while the kernel governs important characteristics such as continuity and smoothness. The requirement for a kernel is that it needs to be positive semidefinite. Intuitively, the kernel defines the similarity between two points based on their mutual distance. In this section, we present the fundamental concepts of GP regression and dimensionality reduction using pseudo-inputs, and construct the surrogate function based on these principles. Moreover, we discuss how to choose the mean and the kernel.

\subsection{GP Regression} \label{sec:gp_pred}

GPs can be adopted to approximate a nonlinear function using samples of the function at known input points. After acquiring the samples, the GP can be evaluated at any input point. The evaluation is done by finding the posterior distribution of the GP at the input point, i.e., conditioning on the observed inputs and outputs. In general, complex-valued GPs require specifying the pseudo-covariance that relates the real and imaginary parts of the signal \cite{boloix2018complex}. However, under the assumption of circular symmetry of the channels and the assumed form of $g$ in~\eqref{eq:lnadistortion}, the real and imaginary parts of the output of the distortion function remain uncorrelated. Consequently, the complex GP prior reduces to two independent real GPs with identical kernels. This enables the use of standard real-valued kernels with complex inputs by treating the real and imaginary components separately but symmetrically. Consider \eqref{eq:systemmodel} and define $\g = g(\z)$. We impose a Gaussian prior $\g \sim \mathcal{CN}(\m_{\g}, \C_{\g})$, where $\m_{\g} \in \Compl^{MN}$ is the mean vector and $\C_{\g} \in \Real^{MN \times MN}$ is the covariance matrix. The mean vector is acquired by sampling the mean function at the inputs, whereas the covariance matrix is obtained by evaluating the kernel pairwise between inputs, i.e.,
\begin{align} \label{eq:mvec}
    \m_{\g} & = 
    \begin{bmatrix}
        \mu(z_1) \\
        \vdots \\
        \mu(z_{MN})
    \end{bmatrix}, \\
    \C_{\g} & =
    \begin{bmatrix} \label{eq:covmat}
        K(z_1, z_1) & \dots & K(z_1, z_{MN}) \\
        \vdots & \ddots & \vdots \\
        K(z_{MN}, z_1) & \dots & K(z_{MN}, z_{MN})
    \end{bmatrix}.
\end{align}
We define $\widehat{g}(\z)$ as the expectation of $\g$ when conditioned on $\y$, i.e.,
\begin{equation}
    \widehat{g}(\z) = \mathbb{E}[\g|\y].
\end{equation}
Then, the joint distribution of $\g$ and $\y$ is
\begin{equation}
    \begin{bmatrix}
        \g \\
        \y
    \end{bmatrix}
    \sim
    \mathcal{CN}
    \left(
    \begin{bmatrix}
        \m_{\g} \\
        \m_{\g}
    \end{bmatrix},
    \begin{bmatrix}
        \C_{\g} & \C_{\g} \\
        \C_{\g} & \C_{\g} + \sigma^2 \I_{MN}
    \end{bmatrix}
    \right).
\end{equation}
Finally, using the properties of multivariate Gaussian distributions, the conditional expectation of $\g$ given $\y$ is expressed~as
\begin{equation} \label{eq:gp_pred}
    \mathbb{E}[\g|\y] = \m_{\g} + \C_{\g} (\C_{\g} + \sigma^2 \I_{MN})^{-1} (\y - \m_{\g}).
\end{equation}

\subsection{Dimensionality Reduction Using Pseudo-Inputs} \label{sec:gp_dimred}

From \eqref{eq:gp_pred}, it can been seen that the prediction requires inverting the dense $MN \times MN$ matrix $\C_{\g\g} + \sigma^2 \I_{MN}$, which is computationally demanding when $M N$ is large. Several approaches have been proposed, such as sparse GPs \cite{snelson2005sparse}, to reduce this computational load. In this work, we exploit a related approach that parametrizes $\g$ using the pseudo-inputs $\widetilde{\z} \in \Compl^D$, with $D \ll MN$. In contrast with \cite{snelson2005sparse}, where the locations of the pseudo-inputs are estimated along with other unknowns, we fix them prior to the estimation. We introduce a random variable $\widetilde{\g} = g(\widetilde{\z}) \in \Compl^D$ such that $\widetilde{\g} \sim \mathcal{CN}(\m_{\widetilde{\g}}, \C_{\widetilde{\g}\widetilde{\g}})$, with
\begin{align} \label{eq:mvectilde}
    \m_{\widetilde{\g}} & = 
    \begin{bmatrix}
        \mu(\widetilde{z}_1) \\
        \vdots \\
        \mu(\widetilde{z}_D)
    \end{bmatrix} \in \Compl^D, \\
    \C_{\widetilde{\g}} & =
    \begin{bmatrix} \label{eq:covmattilde}
        K(\widetilde{z}_1, \widetilde{z}_1) & \dots & K(\widetilde{z}_1, \widetilde{z}_D) \\
        \vdots & \ddots & \vdots \\
        K(\widetilde{z}_D, z_1) & \dots & K(\widetilde{z}_D,\widetilde{z}_D)
    \end{bmatrix} \in \Real^{D \times D}.
\end{align}
Moreover, we define the cross-covariance between $\g$ and $\widetilde{\g}$ as
\begin{equation} \label{eq:crosscovmat}
    \C_{\g\widetilde{\g}} = 
    \begin{bmatrix}
        K(z_1, \widetilde{z}_1) & \dots & K(z_1, \widetilde{z}_D) \\
        \vdots & \ddots & \vdots \\
        K(z_{MN}, \widetilde{z}_1) & \dots & K(z_{MN}, \widetilde{z}_D)
    \end{bmatrix}
    \in \Real^{MN \times D}.
\end{equation}
Thus, the joint distribution of $\g$ and $\widetilde{\g}$ is
\begin{equation}
    \begin{bmatrix}
        \g \\
        \widetilde{\g}
    \end{bmatrix}
    \sim
    \mathcal{CN}
    \left(
    \begin{bmatrix}
        \m_{\g} \\
        \m_{\widetilde{\g}}
    \end{bmatrix},
    \begin{bmatrix}
        \C_{\g} & \C_{\g\widetilde{\g}} \\
        \C_{\g\widetilde{\g}}^\Trans & \C_{\widetilde{\g}}
    \end{bmatrix}
    \right),
\end{equation}
whereas the expectation of $\g$ given $\widetilde{\g}$ is
\begin{equation}
    \mathbb{E}[\g|\widetilde{\g}] = \m_{\g} + \C_{\g\widetilde{\g}} \C_{\widetilde{\g}\widetilde{\g}}^{-1} (\widetilde{\g} - \m_{\widetilde{\g}}).
\end{equation}
We now write \eqref{eq:systemmodel} in terms of $\widetilde{\g}$ as
\begin{equation} \label{eq:systemmodel_dimred}
    \y = \mathbb{E}[\g|\widetilde{\g}] + \e = \m_{\g} + \C_{\g\widetilde{\g}} \C_{\widetilde{\g}\widetilde{\g}}^{-1} (\widetilde{\g} - \m_{\widetilde{\g}}) + \e.
\end{equation}
Again, from the joint distribution of $\widetilde{\g}$ and $\y$, after some matrix manipulations, we obtain
\begin{equation} \label{gtilde_surrogate}
    \mathbb{E}[\widetilde{\g}|\y] = \m_{\widetilde{\g}} + (\B^\Trans \B + \sigma^2 \C_{\widetilde{\g}}^{-1} )^{-1} \B^\Trans (\y - \m_{\g}),
\end{equation}
where we have defined $\B = \C_{\g\widetilde{\g}}\C_{\widetilde{\g}}^{-1} \in \Real^{MN \times D}$. Substituting \eqref{gtilde_surrogate} in place of $\widetilde{\g}$ in \eqref{eq:systemmodel_dimred}, we obtain the lower-dimensional surrogate function
\begin{equation} \label{eq:gp_pred_dimred}
    \widehat{g}_\textrm{r}(\z) = \m_{\g} + \B (\B^\Trans \B + \sigma^2 \C_{\widetilde{\g}}^{-1} )^{-1} \B^\Trans (\y - \m_{\g}).
\end{equation}
Note that here we only need to invert the $D \times D$ matrix $\B^\Trans \B + \sigma^2 \C_{\widetilde{\g}}^{-1}$.

\subsection{Choice of the Mean Function and Kernel} \label{sec:meancov}

The choice of the mean function primarily affects how the GP behaves in regions with no observed inputs. When the distance between the point at which we want to predict the function and the observed inputs increases, the predicted value converges to the value of the mean function at that point. We choose the mean function as $\mu(z) = z$, so the GP effectively models deviations from the linear function $\z = \A \u$. In other words, we assume that the underlying relationship is predominantly linear with comparatively minor nonlinear distortions, which is valid when the nonlinear distortions constitute a small perturbation relative to the main linear structure of the signal. On the other hand, the choice of the kernel affects, for instance, the smoothness properties of the GP. We use the squared exponential kernel
\begin{equation} \label{eq:kernel}
    K(z, z') = \tau^2 \exp \left(-\rho^2 |z - z'|^2 \right)
\end{equation}
between points $z, z' \in \Compl$, with signal variance $\tau^2 \geq 0$ and inverse length-scale $\rho \geq 0$. The signal variance controls the scale of the nonlinear term of the GP and the length-scale controls local fluctuations. Decreasing $\rho$ makes the GP stiffer and less likely to overfit to the AWGN. This kernel is a standard choice in function approximation when no strong assumptions about the shape of the function are made. The kernel results in a GP that is infinitely mean-square real differentiable (and, hence, very smooth). We recall that the input of the GP is $\z = \A\u$, which results in
\begin{equation} \label{eq:covfunc}
    \begin{split}
        K(z_j, z_l) &= \tau^2 \exp\left(-\rho^2|z_j - z_l|^2\right) \\
        &= \tau^2 \exp\left(-\rho^2(\A_{j:} - \A_{l:})\u \u^\Herm(\A_{j:} - \A_{l:})^\Herm\right) \\
        &= \tau^2 \exp\left(-(\A_{j:} - \A_{l:})\u_\rho \u_\rho^\Herm(\A_{j:} - \A_{l:})^\Herm\right).
    \end{split}
\end{equation}
From \eqref{eq:covfunc}, we observe that the inverse length-scale $\rho$ can be absorbed into $\u$, creating scaling ambiguity. In wireless systems, downstream tasks such as detection and beamforming are inherently scale-invariant and thus effectively operate on normalized channel estimates; as a result, only the phases and relative amplitudes of the channel influence the system's performance. Hence, the scale of the result is not meaningful and this ambiguity is not an issue. Moreover, since the kernel depends only on the distance between the two points, it is rotation-invariant, possibly causing unwanted phase shifts in the estimated channel. Therefore, the estimation result must be rotated back to its original axis, e.g., following the procedure proposed in Section~\ref{sec:rotation}.

Note that, in the algorithmic implementation, the GP mean and covariance matrices are not evaluated directly on the noisy received signal but on an estimate of the underlying noiseless quantity $\mathbf{z} = \mathbf{A}\mathbf{u}$. Throughout the inference procedure, the algorithm maintains an estimate of $\mathbf{z}$, which serves as the effective input to the GP kernel. This ensures that the GP is conditioned on a denoised representation of the latent signal.

\section{Estimation of the Sparse Channel} \label{sec:estimation}

In this section, we introduce a hierarchical prior model based on SBL that enforces sparsity of the estimated angular channels. Moreover, we propose two enhanced SBL methods, \textit{nonlinear enhanced SBL (NL-E-SBL)} and \textit{nonlinear modified enhanced SBL (NL-M-E-SBL)}, for estimating the channels when the GP-based surrogate function is used in place of the distortion function $g(\cdot)$ to model the hardware impairments. These methods are extensions of their linear counterparts (E-SBL and M-E-SBL) presented in \cite{arjas2025enhanced}.

\subsection{Hierarchical Prior Model} \label{sec:hierarchicalprior}

To exploit the angular sparsity of the channels, we adopt a Bayesian approach using a hierarchical prior distribution that leads to a computationally efficient estimation of the channels. Each element of the sparse channel $\u$ is assigned a Gaussian prior with unknown element-specific variances. Moreover, the variances are assigned inverse-gamma priors. This construction results in a heavy-tailed Student's $t$ prior distribution, thereby promoting sparsity. Formally, we introduce a weight vector $\w \in \Real^{MK}$ and a scaling vector $\s \in \Real^{MK}$, and set
\begin{align}
    u_j|w_j, s_j &\sim \mathcal{CN}(0, s_j w_j), \label{eq:hierarchicalmodel1} \\ 
    w_j &\sim \mathcal{IG}(\nu/2, \nu/2), \label{eq:hierarchicalmodel2} \\ 
    s_j &\sim \mathcal{IG}(\gamma, \beta), \label{eq:hierarchicalmodel3}
\end{align}
$\forall j = 1, \dots, MK$. The hyperparameters $\nu$, $\gamma$, $\beta > 0$ are fixed prior to the estimation.

In the standard SBL formulation, each coefficient's variance is governed by a single inverse-gamma prior with common shape parameters, which enforces identical tail behavior across the coefficients and thus a uniform level of shrinkage. This can be suboptimal when the signal contains components with widely different magnitudes. The hierarchical prior used in our formulation alleviates this limitation by introducing an extra hyperparameter layer that lets each coefficient adapt its own effective variance and tail profile. As a result, large coefficients are penalized less aggressively while small ones are more strongly suppressed, providing a more flexible and data-driven sparsity structure. The hierarchical SBL prior adopted here treats the elements of \(\u\) as independent for analytical tractability and consistency with the standard SBL framework. While this does not explicitly capture potential spatial correlations in the channel, such correlations could in principle be incorporated through structured or covariance-aware priors as in \cite{prasanna2021mmwave}, at the cost of higher computational complexity.

In the following, we formulate two methods for estimating the model parameters, i.e., NL-E-SBL and NL-M-E-SBL.

\subsection{Nonlinear Enhanced SBL (NL-E-SBL)} \label{sec:NL-E-SBL}

The goal of NL-E-SBL is to compute the \textit{maximum a posteriori estimate of $\w$ and $\s$ after marginalizing over $\u$}, where the iterative expectation-maximization (EM) algorithm \cite{dempster1977maximum} is used to maximize the marginal posterior density. Since the dependence on $\u$ is nonlinear, we re-linearize the function $\widehat{g}_\textrm{r} (\cdot)$ at each iteration so that we can analytically marginalize it. This idea is used, for example, in the extended Kalman filter \cite{sarkka2013bayesian}. This linearization means that the usual monotone convergence properties of the algorithm are not guaranteed without properly chosen step sizes in the updates. As $\widehat{g}_\textrm{r} (\cdot)$ is non-holomorphic, we must utilize Wirtinger derivatives in the linearization.

Let $\u^t \in \Compl^{M K}$, $\w^t \in \Real^{M K}$, and $\s^t \in \Real^{MK}$ denote the estimates of $\u$, $\w$, and $\s$, respectively, at iteration $t$. Moreover, define $\widetilde{\u} = [\u^\Trans,\u^\Herm]^\Trans \in \Compl^{2 M K}$ and $\widetilde{\y} = [\y^\Trans,\y^\Herm]^\Trans \in \Compl^{2 M N}$. The EM algorithm iterates between the expectation step (E-step) and the maximization step (M-step). At iteration $t$, the E-step defines the expected value of the non-marginalized log-posterior density with respect to $\widetilde{\u}$ conditioned on $\widetilde{\y}$ and the current parameter estimates $\w^t$ and $\s^t$, i.e.,
\begin{equation} \label{eq:E-step}
    Q(\w, \s|\w^t, \s^t) = \mathbb{E}_{\widetilde{\u}|\widetilde{\y}, \w^t, \s^t} \big[\log \big(p(\widetilde{\u}|\w, \s) p(\w) p(\s) \big) \big].
\end{equation}
Due to the nonlinearity, the distribution of $\widetilde{\u}|\widetilde{\y}, \w^t, \s^t$ is not analytically tractable. Therefore, following \cite{amin2011wirtinger}, we linearize $\widehat{g}_\textrm{r} (\cdot)$ at $\z^t = \A \u^t$ and obtain the linearized system model
\begin{equation} \label{eq:linmodel}
    \begin{split}
    \widetilde{\y} 
    & = \begin{bmatrix}
        \widehat{g}_\textrm{r}(\z^t) \\
        \widehat{g}_\textrm{r}(\z^t)^*
    \end{bmatrix}
    + 
   \widetilde{\J}_{\z^t}
    \begin{bmatrix}
        \u - \u^t \\
        \u^* - \u^{t*}
    \end{bmatrix}
    +
    \begin{bmatrix}
        \e \\
        \e^*
    \end{bmatrix} + \mathcal{O}\big(\|\u - \u^t\|^2\big),
    \end{split}
\end{equation}
with
\begin{equation} \label{eq:full_jacobian}
    \widetilde{\J}_{\z^t} = 
    \begin{bmatrix}
        \J_{\z^t} \A & \J_{\z^{t*}} \A^* \\
        \J_{\z^{t*}}^* \A & \J_{\z^t}^* \A^*
    \end{bmatrix}.
\end{equation}
Here, $\J_{\z^t} \in \Compl^{MN \times MN}$ and $\J_{\z^{t*}} \in \Compl^{MN \times MN}$ denote, respectively, the Jacobian and conjugate Jacobian matrices of $\widehat{g}_\textrm{r} (\cdot)$ evaluated at $\z^t$. We refer to Section~\ref{sec:jacobians} for their definitions and computation details. If we omit the remainder $\mathcal{O}\big(\|\u - \u^t\|^2\big)$ in \eqref{eq:linmodel}, we have $\widetilde{\y} \sim~\mathcal{CN}(\m_{\widetilde{\y}}^t, \C_{\widetilde{\y}}^t)$, with
\begin{align}
    \m_{\widetilde{\y}}^t & = 
    \begin{bmatrix}
        \widehat{g}_\textrm{r}(\z^t) \\
        \widehat{g}_\textrm{r}(\z^t)^*
    \end{bmatrix}
    -
    \widetilde{\J}_{\z^t}
    \begin{bmatrix}
        \u^t \\
        \u^{t*}
    \end{bmatrix}, \\
    \C_{\widetilde{\y}}^t & = \widetilde{\J}_{\z^t} \widetilde{\W}^t \widetilde{\J}_{\z^t}^\Herm + \sigma^2 \I_{2MN},
\end{align}
with $\widetilde{\W}^t = \begin{bmatrix}
    \W^t & \0 \\
    \0 & \W^t
\end{bmatrix}$ and  $\W^t = \Diag(\w^t \odot \s^t)$. Due to the linearization, the distribution of $\widetilde{\u}|\widetilde{\y}, \w^t, \s^t$ is Gaussian, i.e.,
\begin{equation}
    \widetilde{\u}|\widetilde{\y}, \w^t, \s^t \sim \mathcal{CN}(\mub^t, \Sigmab^t),
\end{equation}
with 
\begin{align} \label{eq:mvec_covmat}
    \mub^t &= \frac{1}{\sigma^2} \Sigmab^t \widetilde{\J}_{\z^t}^\Herm(\widetilde{\y} - \m_{\widetilde{\y}}^t), \\
    \Sigmab^t &= \left(\frac{1}{\sigma^2} \widetilde{\J}_{\z^t}^\Herm \widetilde{\J}_{\z^t} + (\widetilde{\W}^t)^{-1}\right)^{-1}.
\end{align}
The expression \eqref{eq:E-step} in the E-step is thus given by
\begin{equation}
    \begin{split}
        & Q(\w, \s|\w^t, \s^t) \\
        &=-\log \det \widetilde{\W} - (\mub^t)^\Herm \widetilde{\W}^{-1} \mub^t - \trace(\widetilde{\W}^{-1} \Sigmab^t) \\
        &\phantom{=} ~-\frac{\nu + 2}{2}\sum_{j=1}^{MK} \log(w_j) - \frac{\nu}{2} \sum_{j=1}^{MK}\frac{1}{w_j} \\
        &\phantom{=} ~-(\gamma + 1)\sum_{j=1}^{MK} \log(s_j) - \beta \sum_{j=1}^{MK}\frac{1}{s_j}.
    \end{split}
\end{equation}
In the M-step, we differentiate this expression separately with respect to the element of $\w$ and $\s$ while keeping the rest of the elements fixed, and set the derivatives to zero to obtain the update formulas
\begin{align}
    w_j^{t + 1} &= \frac{\nu/2 + 2 \big(|\mu_j^t|^2 + \Sigma_{jj}^t \big)/s_j^t}{\nu/2 + 3}, \label{eq:w_update_NL-E-SBL} \\
    s_j^{t + 1} &= \frac{\beta + 2 \big(|\mu_j^t|^2 + \Sigma_{jj}^t \big)/w_j^{t+1}}{\gamma + 3}, \label{eq:s_update_NL-E-SBL}
\end{align}
$\forall j = 1, \dots, MK$. After updating $\w$ and $\s$, we set $\widetilde{\u}^{t+1} = \delta \mub^t + (1 - \delta)\widetilde{\u}^t$, where $\delta \in (0, 1]$ is a step size that prevents the algorithm from diverging. This procedure is repeated until convergence.

\subsection{Nonlinear Modified E-SBL (NL-M-E-SBL)} \label{sec:NL-M-E-SBL}

To update $\w$ and $\s$ in NL-E-SBL, one has to compute the diagonal values of the matrix $\Sigmab^t$, which is computationally demanding when $MK$ is large. NL-M-E-SBL aims to bypass these computations, which results in a more computationally efficient algorithm and slightly different estimates. In contrast to NL-E-SBL, which finds the maximum a posteriori estimate of $\w$ and $\s$ after marginalizing over $\u$, the goal of NL-M-E-SBL is to compute the \textit{maximum a posteriori estimate of the joint distribution of $\u$, $\w$, and $\s$}. By Bayes' formula, the posterior density of $(\widetilde{\u}, \w, \s)$ is
\begin{equation} \label{eq:joint_posterior_density}
    \begin{split}
        p(\widetilde{\u}, \w, \s| \y) & \propto p(\y|\widetilde{\u})p(\widetilde{\u}|\w, \s)p(\w)p(\s) \\
        &= \exp\left\{-\frac{1}{\sigma^2} \|\y - \widehat{g}_\textrm{r}(\A\u)\|^2 \right\} \\
        &\phantom{=} ~ \times \frac{1}{\det(\widetilde{\W})} \exp\big\{- \widetilde{\u}^\Herm \widetilde{\W}^{-1} \widetilde{\u} \big\} \\
        &\phantom{=} ~ \times \prod_{j=1}^{MK} w_j^{-\frac{\nu + 2}{2}} \exp \left\{-\frac{\nu}{2 w_j}\right\} \\
        &\phantom{=} ~ \times \prod_{j=1}^{MK} s_j^{-(\gamma + 1)} \exp \left\{ -\frac{\beta}{s_j} \right\}.
    \end{split}    
\end{equation}
For convenience, we take the logarithm of \eqref{eq:joint_posterior_density} and obtain the objective function
\begin{equation} \label{eq:obj_func}
    \begin{split}
        f(\widetilde{\u}, \w, \s) &= -\frac{1}{\sigma^2}\big\|\widetilde \y - \widetilde{\widehat{g}_\textrm{r}(\A\u)}\big\|^2 \\
        &\phantom{=} ~-\log \det \widetilde{\W} - \widetilde{\u}^\Herm \widetilde{\W}^{-1} \widetilde{\u} \\
        &\phantom{=} ~-\frac{\nu + 2}{2}\sum_{j=1}^{MK} \log(w_j) - \frac{\nu}{2} \sum_{j=1}^{MK}\frac{1}{w_j} \\
        &\phantom{=} ~-(\gamma + 1)\sum_{j=1}^{MK} \log(s_j) - \beta \sum_{j=1}^{MK}\frac{1}{s_j},
    \end{split}
\end{equation}
with $\widetilde{\widehat{g}_\textrm{r}(\A \u)} = [\widehat{g}_\textrm{r}(\A \u)^\Trans \ \widehat{g}_\textrm{r}(\A \u)^\Herm]^\Trans$. In general, the above objective function is nonconvex, making it challenging to reach a global optimum. However, we can achieve a local optimum quite efficiently by optimizing each parameter vector, i.e., $\widetilde{\u}$, $\w$, and $\s$, in an alternating fashion while keeping the other two fixed. Due to the presence of the nonlinear function $\widehat{g}_\textrm{r} (\cdot)$, the optimization of $\widetilde{\u}$ has to be performed iteratively, whereas the updates of $\w$ and $\s$ can be derived in closed form.

The update of $\widetilde{\u}$ involves maximizing the objective function
\begin{equation} \label{eq:obj_func_u}
    f_\u(\widetilde{\u}) = -\frac{1}{\sigma^2}\big\|\widetilde \y - \widetilde{\widehat{g}_\textrm{r}(\A\u)}\big\|^2 - \widetilde{\u}^\Herm \widetilde{\W}^{-1} \widetilde{\u}.
\end{equation}
To find the maximizer, we utilize the Gauss-Newton (GN) method, which is based on the successive linearization and optimization of the resulting quadratic function. The complex valued version of the GN method that utilizes Wirtinger calculus is described in \cite{amin2011wirtinger}. The GN update formula at iteration $t$ is given by
\begin{equation} \label{eq:u_update_NL-M-E-SBL}
    \widetilde{\u}^{t+1} = \frac{\delta}{\sigma^2} \Sigmab^t \widetilde{\J}_{\z^t}^\Herm(\widetilde{\y} - \m_{\widetilde{\y}}^t) + (1 - \delta)\widetilde{\u}^t.
\end{equation}

After rearranging \eqref{eq:obj_func}, the update of $\w$ involves maximizing the objective function
\begin{equation}
    f_\w(\w) = -\! \sum_{j = 1}^{MK} \bigg( \! 2\log(w_j s_j) + 2\frac{|u_j|^2}{w_j s_j} + \frac{\nu + 2}{2}\log(w_j) + \frac{\nu}{2w_j} \! \bigg),
\end{equation}
which gives the update formula
\begin{equation} \label{eq:w_update_NL-M-E-SBL}
    w_j^{t+1} = \frac{\nu/2 + 2|u_j^t|^2/s_j^t}{\nu/2 + 3}, ~ \forall j = 1, \dots, MK.
\end{equation}
Lastly, the update of $\s$ involves maximizing the objective function
\begin{equation}
    f_\s(\s) = -\sum_{j = 1}^{MK} \bigg(2\log(w_j s_j) + 2\frac{|u_j|^2}{w_j s_j} + (\gamma + 1)\log(s_j) + \frac{\beta}{s_j}\bigg),
\end{equation}
which gives the update formula
\begin{equation} \label{eq:s_update_NL-M-E-SBL}
    s_j^{t+1} = \frac{\beta + 2|u_j^t|^2/w_j^{t+1}}{\gamma + 3}, ~ \forall j = 1, \dots, MK.
\end{equation}
Each parameter vector is updated in an alternating fashion while keeping the other two fixed, and this procedure is repeated until convergence.

\subsection{Differences Between NL-E-SBL and NL-M-E-SBL} \label{sec:differences}

Although NL-E-SBL and NL-M-E-SBL are based on the same underlying model, their update rules have different theoretical guarantees. In NL-E-SBL, the linearization inside the E-step does not produce a valid lower bound on the marginal likelihood: therefore, the monotone ascent property associated with classical EM does not hold, and NL-E-SBL should be regarded as a practical heuristic extension of E-SBL that renders the otherwise intractable E-step computationally feasible. In contrast, NL-M-E-SBL is constructed to ensure monotonicity of the objective: each update of $\u$ increases the objective for a sufficiently small step size $\delta$, and the updates of $\w$ and $\s$ become strictly concave maximization problems after a log-reparameterization, ensuring unique solutions that further increase the objective. A concise convergence proof for NL-M-E-SBL is provided in the Appendix.

When the posterior distribution is sharply concentrated, both approaches produce nearly identical estimates, which explain their similar performance in the numerical results. Noticeable differences emerge primarily in settings with weaker sparsity, such as those with fewer antennas, where the posterior becomes less concentrated.

\subsection{Rotating the Result Back to the Original Axis} \label{sec:rotation}

As mentioned in Section~\ref{sec:meancov}, the squared exponential kernel in \eqref{eq:covfunc} for complex inputs is rotation-invariant, i.e., its output does not change when both its inputs are rotated by the same angle. This might result in convergence issues and unwanted phase shifts in the estimation results. While this problem is partially resolved by adding the mean function $\mu(z) = z$, which is not rotation-invariant, the estimated channels might still be slightly phase shifted. Notably, we can undo the phase shift by simple post-processing, where we rotate the channels back to the original axis. The optimal angle $\theta^\star \in [0, 2\pi)$ by which we need to rotate the estimated angular channel $\widehat{\u} \in \Compl^{MK}$ is given by
\begin{equation}
    \begin{split}
        \theta^\star &= \argmin_\theta \|\y - e^{i\theta}\A \widehat{\u}\|^2 \\
        &= \argmin_\theta \big(\|\y\|^2 - 2\Re(e^{-i\theta}\widehat{\u}^\Herm \A^\Herm \y) + \|e^{i\theta}\A \widehat{\u}\|^2 \big) \\
        &= \argmin_\theta \big(-\Re(e^{-i\theta} \widehat{\u}^\Herm \A^\Herm \y) \big).
    \end{split}
\end{equation}
Let us derive the real part of $e^{-i\theta} \widehat{\u}^\Herm \A^\Herm \y$. We denote $w = \widehat{\u}^\Herm \A^\Herm \y$ and apply Euler's formula as
\begin{equation}
    \begin{split}   
        e^{-i\theta} w &= (\cos \theta - i \sin \theta)\big(\Re(w) + i\Im(w)\big) \\
        &= \Re(w) \cos \theta + \Im(w) \sin \theta \\
        &\phantom{=} ~ + i \big(\Im(w) \cos \theta - \Re(w) \sin \theta \big),
    \end{split}
\end{equation}
which yields $-\Re \big(e^{-i\theta}w \big) = -\Re(w) \cos \theta - \Im(w) \sin \theta$. Differentiating this expression with respect to $\theta$ and setting the derivative to zero provides the optimal angle
\begin{equation} \label{eq:opt_angle}
    \begin{split}
    \frac{\partial}{\partial \theta} (- \Re(e^{-i\theta})) &= \Re(w) \sin \theta - \Im(w) \cos \theta = 0 \\
    \iff \theta^\star &= \tan^{-1} \left(\frac{\Im(w)}{\Re(w)}\right).
    \end{split}
\end{equation}
This formula gives a stationary point that is either a maximizer or a minimizer. To determine which one it is, we evaluate the second derivative
\begin{equation}
    \frac{\partial^2}{\partial \theta^2} \big(-\Re(e^{-i\theta}w) \big) = \Re(w) \cos \theta + \Im(w) \sin \theta
\end{equation}
at $\theta^\star$: if the result is positive, $\theta^\star$ is a minimizer; otherwise, it is a maximizer. In the latter case, we set $\theta^\star \leftarrow~\theta^\star + \pi$, which yields the minimizer. The final result is thus given as $\widehat{\u}_{\theta^\star} = e^{i\theta^\star}\widehat{\u}$.

We note that one could alternatively define the rotation as $\theta^\star = \argmin_\theta \|\y - \widehat{g}_{\mathrm{r}}(e^{i\theta}\A \widehat{\u})\|$. However, since $\widehat{\u}$ is obtained by fitting the GP output to the measurements $\y$, this rotation would be ineffective, as $\widehat{g}_{\mathrm{r}}(\A \widehat{\u})$ is already approximately aligned with $\y$. Instead, we must rotate $\widehat{\u}$ so that the linear prediction $\A \widehat{\u}$ aligns with the measurements. This resolves the global phase ambiguity in $\widehat{\u}$ that remains after the nonlinear fit. This represents the final step of the proposed nonlinear estimation framework, which is outlined in Algorithm~\ref{alg:NL-SBL}.

\begin{algorithm}[t!]
\footnotesize
\caption{NL-E-SBL/NL-M-E-SBL} \label{alg:NL-SBL}
\begin{algorithmic}[1]
\State \textbf{Input:} $\y$, $\A$, $\sigma^2$, $\tau^2$, $\rho^2$, $\nu$, $\gamma$, $\beta$
\State \textbf{Initialize:} $\u^0$, $\w^0$, $\s^0$, $\delta$
\State \textbf{Set:} $\widetilde{\u}^0 = [(\u^0)^\Trans, (\u^0)^\Herm]^\Trans$, $t = 0$
\Repeat
    \State Compute $\z^t = \A \u^t$
    \State Compute Jacobians $\J_{\z^t}$ and $\J_{\z^{t*}}$ as in Section~\ref{sec:jacobians}
    \vspace{3mm}
    \State \textbf{NL-E-SBL}:    
    \State \quad \quad Compute $\mub^t$ and $\Sigmab^t$ as in \eqref{eq:mvec_covmat}
    \State \quad \quad Update $\w$ as in \eqref{eq:w_update_NL-E-SBL}
    \State \quad \quad Update $\s$ as in \eqref{eq:s_update_NL-E-SBL}
    \State \quad \quad Set $\widetilde{\u}^{t+1} \leftarrow \delta \mub^t + (1 - \delta) \widetilde{\u}^t$
    \vspace{3mm}
    \State \textbf{NL-M-E-SBL}:   
    \State \quad \quad Update $\widetilde{\u}$ as in \eqref{eq:u_update_NL-M-E-SBL}
    \State \quad \quad Update $\w$ as in \eqref{eq:w_update_NL-M-E-SBL}
    \State \quad \quad Update $\s$ as in \eqref{eq:s_update_NL-M-E-SBL}
    \vspace{3mm}
    \State Set $\u^{t+1} \leftarrow \widetilde{\u}^{t+1}_{1:MK}$
    \State Set $t \leftarrow t + 1$
\Until{Convergence}
\State Find rotation angle $\theta^\star$ as in \eqref{eq:opt_angle}
\State \textbf{Output:} $e^{i\theta^\star} \u^{t}$
\end{algorithmic}
\end{algorithm}

\subsection{Computation of the Jacobians} \label{sec:jacobians}

In this section, we provide details on how to compute the Jacobians that arise from linearizing the surrogate function $\widehat{g}_\textrm{r} (\cdot)$ in \eqref{eq:linmodel}. In the computation, we utilize Wirtinger calculus, which can be used to optimize real-valued functions with complex inputs. We refer to \cite{sorber2012unconstrained} for an introduction to Wirtinger calculus. The Wirtinger derivatives for a complex number $z = x + iy$ are defined as the differential operators
\begin{equation}
    \frac{\partial}{\partial z} = \frac{1}{2}\left(\frac{\partial}{\partial x} - i\frac{\partial}{\partial y}\right), \quad \frac{\partial}{\partial z^*} = \frac{1}{2}\left(\frac{\partial}{\partial x} + i\frac{\partial}{\partial y}\right).
\end{equation}
These definitions can be extended to the multivariate setting. Considering \eqref{eq:linmodel}, the Jacobian $\J_\z$ and conjugate Jacobian $\J_{\z^*}$ at $\z$ are defined as
\begin{align}
    \J_\z &= \begin{bmatrix} \label{eq:Jz}
        \displaystyle \frac{\partial [\widehat{g}
        _\textrm{r}(\z)]_1}{\partial z_1} & \dots & \displaystyle \frac{\partial [\widehat{g}_\textrm{r}(\z)]_1}{\partial z_{MN}} \\
        \vdots & \ddots & \vdots \\
        \displaystyle \frac{\partial [\widehat{g}_\textrm{r}(\z)]_{MN}}{\partial z_1} & \dots & \displaystyle \frac{\partial [\widehat{g}_\textrm{r}(\z)]_{MN}}{\partial z_{MN}}
    \end{bmatrix}, \\
    \J_{\z^*} &= \begin{bmatrix} \label{eq:Jzs}
        \displaystyle \frac{\partial [\widehat{g}_\textrm{r}(\z)]_1}{\partial z^*_1} & \dots & \displaystyle \frac{\partial [\widehat{g}_\textrm{r}(\z)]_1}{\partial z^*_{MN}} \\
        \vdots & \ddots & \vdots \\
        \displaystyle \frac{\partial [\widehat{g}_\textrm{r}(\z)]_{MN}}{\partial z^*_1} & \dots & \displaystyle \frac{\partial [\widehat{g}_\textrm{r}(\z)]_{MN}}{\partial z^*_{MN}}
    \end{bmatrix},
\end{align}
respectively. Recalling the definition of $\widehat{g}_\textrm{r} (\cdot)$ in \eqref{eq:gp_pred_dimred}, we have
\begin{equation}
    \big[\widehat{g}_\textrm{r}(\z)\big]_j = z_j + \B_{j:} (\B^\Trans \B + \sigma^2 \C_{\widetilde{\g}}^{-1})^{-1} \B^\Trans (\y - \z),
\end{equation}
$\forall j = 1, \dots, MN$. Applying the product rule and the fact that $\frac{\partial \Q^{-1}}{\partial \theta} = -\Q^{-1} \left(\frac{\partial \Q}{\partial \theta}\right) \Q^{-1}$ for any invertible matrix $\Q$ gives
\begin{align}
    \begin{split}
    \frac{\partial \big[\widehat{g}_\textrm{r}(\z)\big]_j}{\partial z_l} &= \delta_{jl} + \frac{\partial \B_{j:}}{\partial z_l} \V^{-1} \B^\Trans (\y - \z) \\
    &\phantom{=} ~ + \B_{j:} \V^{-1} \frac{\partial \B^\Trans \B}{\partial z_l} \V^{-1} \B^\Trans (\y - \z) \\
    &\phantom{=} ~ + \B_{j:} \V^{-1} \frac{\partial \B^\Trans}{\partial z_l} (\y - \z) - [\B_{j:} \V^{-1} \B^\Trans]_{l},
    \end{split} \\
    \begin{split}
    \frac{\partial \big[\widehat{g}_\textrm{r}(\z)\big]_j}{\partial z_l^*} &= \frac{\partial \B_{j:}}{\partial z_l^*} \V^{-1} \B^\Trans (\y - \z) \\
    &\phantom{=} ~ + \B_{j:} \V^{-1} \frac{\partial \B^\Trans \B}{\partial z_l^*} \V^{-1} \B^\Trans (\y - \z) \\
    &\phantom{=} ~ + \B_{j:} \V^{-1} \frac{\partial \B^\Trans}{\partial z_l^*} (\y - \z),
    \end{split}
\end{align}
with $\V = \B^\Trans \B + \sigma^2 \C_{\widetilde{\g}}^{-1} \in \Real^{D \times D}$ and where $\delta_{jl}$ denotes the Kronecker delta. Moreover, we have
\begin{equation}
    \frac{\partial \B}{\partial z_j} = \frac{\partial \C_{\g\widetilde{\g}}}{\partial z_j}\C_{\widetilde{\g}}^{-1} \in \Real^{MN \times D}.
\end{equation}
Lastly, the Wirtinger derivatives of the elements of $\C_{\g\widetilde{\g}}$ are given by
\begin{align}
        \frac{\partial [\C_{\g\widetilde{\g}}]_{jl}}{\partial z_j} &= \rho^2 [\C_{\g\widetilde{\g}}]_{jl}(z_j^* - \widetilde{z}_l^*), ~ \forall j,l = 1, \dots, MN,\\
        \frac{\partial [\C_{\g\widetilde{\g}}]_{jl}}{\partial z_j^*} &= \rho^2 [\C_{\g\widetilde{\g}}]_{ji}(z_j - \widetilde{z}_l), ~ \forall j,l = 1, \dots, MN,\\
        \frac{\partial [\C_{\g\widetilde{\g}}]_{jl}}{\partial z_k} &= \frac{\partial [\C_{\g\widetilde{\g}}]_{jl}}{\partial z_k^*} = 0, ~ \forall k \neq j.
\end{align}
Note that the pseudo-inputs $\widetilde{\z}$ are fixed throughout the algorithm, while the inputs $\z^{t} = \A\u^{t}$ are recomputed at each iteration, ensuring that $\m_{\widetilde \g}$, $\C_{\widetilde \g}$, and $\C_{\g \widetilde \g}$, can be constructed directly from known quantities.

\section{Computational Complexity and Implementation} \label{sec:implementation}

In this section, we discuss the computational complexity and implementation aspects of the proposed nonlinear estimation framework.

\subsection{Computational Complexity} \label{sec:implementation_1}

The computational cost of the enhanced SBL methods primarily arises from solving a linear system involving the matrix $\frac{1}{\sigma^2}\widetilde{\J}_{\z^{i}}^\Herm \widetilde{\J}_{\z^{i}} + \widetilde{\W}^{-1} \in \Compl^{2MK \times 2MK}$ at each iteration, which has complexity $\mathcal{O}(8M^3K^3)$. Additionally, in NL-E-SBL, we need to compute the diagonal elements of the inverse of this matrix, matching the cost of solving the linear system. Evaluating the surrogate function $\widehat{g}_\textrm{r} (\cdot)$ has computational complexity $\mathcal{O}(MND^3)$. The computation of the elements of the Jacobians benefits from the sparsity of $\frac{\partial \C_{\g\widetilde{\g}}}{\partial z_j}$, which has nonzero elements only in the $j$th row. Leveraging this structure, the overall complexity of computing the Jacobians is reduced to $\mathcal{O}(2MND^2)$, given that the inverse matrix $\V^{-1}$ is precomputed when evaluating the surrogate function. The dominant contribution to the cost stems from the term involving the derivative of $\B^\Trans \B$, while the remaining terms are less expensive due to the sparse nature of the derivatives and the fact that some of the terms are common to each element of the matrices.

While the number of pseudo-inputs $D$ can be chosen to balance estimation accuracy and computational complexity (as discussed in Section~\ref{sec:implementation_2}), the computational complexity of the proposed methods still scales cubically with the number of antennas. To address this, future work will explore strategies for further reducing the computational complexity, for example, by approximating the Jacobians via low-rank matrix factorizations or by employing deep unfolding approaches, where a neural network is trained to efficiently optimize the objective functions \cite{andrychowicz2016learning}.

\subsection{Choice of the Pseudo-Inputs} \label{sec:implementation_2}

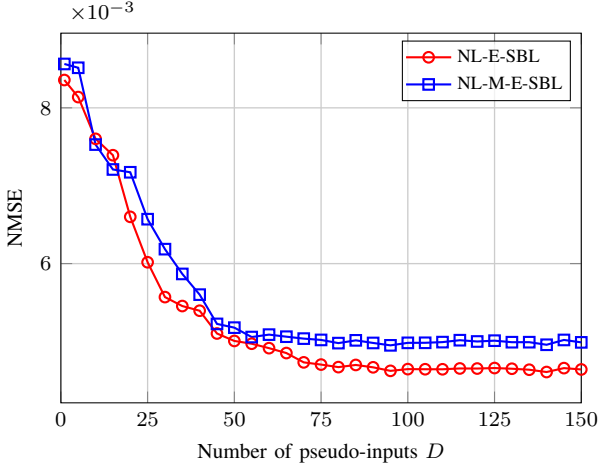
\begin{figure}[!t]
    \centering
    \begin{tikzpicture}

\begin{axis}[
	width=8.5cm,
	height=6.5cm,
	xmin=0, xmax=150,
	xlabel={Number of pseudo-inputs $D$},
	ylabel={NMSE},
	xtick={0,25,...,150},
    tick scale binop=\times,
    label style={font=\footnotesize},
    ticklabel style={font=\footnotesize},
	legend style={at={(0.98,0.98)}, anchor=north east},
	legend style={font=\scriptsize, inner sep=1pt, fill opacity=0.75, draw opacity=1, text opacity=1},
	legend cell align=left,
	grid=both,
	major grid style={line width=0.5pt, draw=gray!40},
    grid style={line width=.2pt, draw=gray!20},
	title={},
	title style={font=\scriptsize, yshift=-2mm},
]

\addplot[thick, red, mark=o]
table [x=Var1, y=NMSEs_E_SBL_GP, col sep=comma] {figures_new/files_txt_new/MSE_D_new.txt};
\addlegendentry{NL-E-SBL};

\addplot[thick, blue, mark=square]
table [x=Var1, y=NMSEs_M_E_SBL_GP, col sep=comma] {figures_new/files_txt_new/MSE_D_new.txt};
\addlegendentry{NL-M-E-SBL};

\end{axis}

\end{tikzpicture}
    \caption{NMSE versus number of pseudo-inputs, with $M = 128$, $K = 5$, $N = 19$, and $\textrm{SNR} = 12$~dB.}
    \label{fig:MSE_D}
\end{figure}

Before running the algorithms, the number and locations of the pseudo-inputs must be specified. For a given number of pseudo-inputs $D$, we determine the locations using Sobol sequences \cite{sobol1967distribution}, which are low-discrepancy sequences filling a space in a highly uniform manner. Uniform coverage is important because the squared exponential kernel in \eqref{eq:kernel} depends on pairwise distances: if the pseudo-inputs cluster locally, their pairwise kernel values become nearly identical, leading to a poorly conditioned kernel matrix $\C_{\widetilde{\g}}$ and potential numerical instability. Unlike purely random sampling, Sobol sequences minimize such clustering while allowing an arbitrary number of pseudo-inputs, offering a practical alternative to rigid grid designs. Specifically, we use a two-dimensional Sobol sequence to represent the real and imaginary parts of the pseudo-inputs. In our numerical results, we scale the signal variance to be approximately one and map the Sobol sequence onto a plane bounded by $-4$ and $4$ in both the horizontal and vertical directions, which ensures that the signal remains within the boundaries with high probability. We note that it is preferable for the boundaries to be overly loose rather than overly strict, as enlarging the area does not degrade the prediction performance, whereas narrowing it might. To choose the number of pseudo-inputs, we study the performance of the proposed methods with respect to $D$. In this regard, 

Fig.~\ref{fig:MSE_D} plots the NMSE of the channel estimation, defined as $\textrm{NMSE} = \frac{\mathbb{E}[\|\widehat{\H} - \H\|^2_\mathrm{F}]}{\mathbb{E}[\|\H\|^2_\mathrm{F}]}$, where $\widehat{\H}$ denotes the estimated channel matrix and $\H$ is the ground truth, as a function of the number of pseudo-inputs. The expectation is computed by averaging over 2000 independent channel realizations. For both NL-E-SBL and NL-M-E-SBL, the NMSE decreases with $D$ up to approximately $D = 75$, after which it plateaus: this demonstrates that substantial dimensionality reduction can be achieved without compromising the estimation accuracy. We note that the precise behavior may also depend on the specific choice of system parameters.

\subsection{Step Size Adaptation and Initialization} \label{sec:stepsize}

The proposed enhanced SBL methods incorporate a step size to prevent the algorithms from diverging. This step size is chosen by means of backtracking line search: if the objective function \eqref{eq:obj_func_u} does not decrease for the new iterate $\u^{t+1}$, the step size is reduced by a factor of $\frac{1}{2}$. The initialization of $\u^0$ is also critical, especially when the magnitude of the distortion terms increases. This causes the suboptimal local optimizers in the objective function to be more prominent, increasing the risk of converging to one of them. To address this, we employ a heuristic initialization that has proven effective in our numerical results. Specifically, we compute the least-squares (LS) estimate $\u_\textrm{LS} = (\A^\Herm \A)^{-1}\A^\Herm \y$ and set all the elements with modulus smaller than $\frac{1}{2}$ to zero. This procedure retains the angles from which most of the signal is received by the BS, thereby promoting convergence to a more desirable solution.

\section{Special Cases} \label{sec:special_cases}

Hardware cost, complexity, and power efficiency are key design considerations in practical wireless systems. These can be addressed, for instance, by employing hybrid analog-digital architectures or fully digital architectures with low-resolution ADCs \cite{heath2016overview}. The first approach introduces an analog beamforming stage to reduce the number of radio frequency (RF) chains \cite{molisch2017hybrid,ahmed2018survey}, whereas the second decreases the resolution of the ADCs (even down to 1 bit) while keeping one RF chain per antenna \cite{Atz21b,Atz22}. In this section, we adapt the proposed methods to accommodate hybrid analog-digital beamforming and 1-bit ADCs.

\subsection{Hybrid Analog-Digital Beamforming} \label{sec:special_cases_hb}

Assuming hybrid analog-digital beamforming, we introduce the analog combiner $\Q \in \Compl^{M\times M_\textrm{RF}}$, where $M_\textrm{RF}$ is the number of RF chains at the BS. In this setting, the received signal in \eqref{eq:systemmodel} becomes
\begin{equation}
    \y_{\textrm{hb}} =\widetilde{\Q}^\Herm \big( g(\z) + \e \big) \in \Compl^{M_\textrm{RF} N},
\end{equation}
with $\widetilde{\Q} = \I_{N} \otimes \Q \in \Compl^{MN \times M_\textrm{RF}N}$. Following the same logic as in Sections \ref{sec:gp_pred} and \ref{sec:gp_dimred}, we obtain the GP-based surrogate function
\begin{equation}
\begin{split}
    \widehat{g}_\textrm{hb}(\z) & = \widetilde{\Q}^\Herm \m_{\g} + \B_{\widetilde{\Q}}(\B_{\widetilde{\Q}}^\Herm \B_{\widetilde{\Q}} + \sigma^2 \C_{\widetilde{\g}\widetilde{\g}}^{-1})^{-1} \\
    & \phantom{=} ~ \times \B_{\widetilde{\Q}}^\Herm(\y_\textrm{hb} - \widetilde{\Q}^\Herm \m_{\g}),
\end{split}
\end{equation}
with $\B_{\widetilde{\Q}} = \widetilde{\Q}^\Herm \B$. This surrogate function can be readily used in place of $\widehat{g}_\textrm{r} (\cdot)$ in NL-E-SBL and NL-M-E-SBL. This has a only a minor impact on the Jacobian computations, while all the other aspects of the methods remain unchanged.

\subsection{1-Bit ADCs} \label{sec:special_cases_1bit}

1-bit ADCs quantize the real and imaginary parts of the received signal to $\pm 1$ (possibly with some scaling). In this setting, the observed signal after the 1-bit ADCs is\footnote{With 1-bit ADCs, unequal received powers across the users can severely degrade the estimation accuracy of the weak users. This effect is typically avoided in the 1-bit massive MIMO literature by assuming uplink power control; see, e.g.,~\cite{li2017channel,Din25,Atz21b,Atz22}. In our numerical results, we follow this convention by assuming identical large-scale fading across the users.}
\begin{equation}
    \r = \sgn \big(\mathrm{Re}(\y)\big) + i \, \sgn \big(\mathrm{Im}(\y)\big) \in \Compl^{MN},
\end{equation}
with $\y$ in \eqref{eq:systemmodel}. We emphasize that the 1-bit quantization occurs only after the signal has passed through the LNAs, and the quantizer is not modeled as part of the GP. Instead of a Gaussian likelihood, we employ a likelihood function appropriate for binary observations, while retaining the GP model solely for the LNA distortion. After minor modifications, the proposed nonlinear estimation framework can be applied to this case as well. The idea is to write the likelihood function induced by the 1-bit ADCs and sequentially approximate it with quadratic functions. The approximations can be interpreted as a Gaussian model with independent error terms characterized by individual variances. Using this Gaussian model, we can define the surrogate function and utilize either NL-E-SBL or NL-M-E-SBL to estimate the channels. This is an example of sequential quadratic programming \cite{nocedal1999numerical}. Although the probit likelihood, which is based on the Gaussian cumulative distribution function (CDF) and used in e.g., \cite{choi2016near}, is the correct model for binary observations arising from a noisy sign process, we adopt the logistic likelihood as a numerically stable and widely used alternative. At high SNRs, the Gaussian CDF becomes increasingly steep and approaches a step function, which can cause numerical instability during optimization. For convenience, we first separate the real and imaginary parts of $\r$ and $\y$ as $\bar{\r} = [\mathrm{Re}(\r)^\Trans ~ \mathrm{Im}(\r)^\Trans]^\Trans \in \Real^{2MN}$ and $\bar{\y} = [\mathrm{Re}(\y)^\Trans ~ \mathrm{Im}(\y)^\Trans]^\Trans \in \Real^{2MN}$, respectively. The log-likelihood function can thus be written as
\begin{equation} \label{eq:loglike_1bit}
    \log p(\bar{\r}|\bar{\y}) = \sum_{j=1}^{2MN} \log \phi(\bar{r}_j \bar{y}_j)
\end{equation}
where $\phi(t) = \frac{1}{1 + e^{-t}}$ is the sigmoid function. We then form the quadratic approximation of the likelihood function around $\Bar{\y}' \in \Real^{2MN}$ as
\begin{equation}
    \begin{split}
    p_\textrm{Quad}(\bar{\r}|\bar{\y};\bar{\y}') &= p(\bar{\r}|\bar{\y}') + (\bar{\y} - \bar{\y}')^\Trans \nabla p(\bar{\r}|\bar{\y}') \\
    &\phantom{=} ~ - \frac{1}{2} (\bar{\y} - \bar{\y}')^\Trans \bar{\C}_p (\bar{\y} - \bar{\y}'),
    \end{split}
\end{equation}
where $\nabla$ is the gradient and $\bar{\C}_p \in \Real^{2MN \times 2MN}$ is the negative Hessian of $\log p$ at $\bar{\y}'$. We note that the Hessian is diagonal because the likelihood function is expressed as a sum of elementwise terms. Completing the square yields
\begin{equation}
    \begin{split}
    p_\textrm{Quad}(\bar{\r}|\bar{\y};\bar{\y}') = -\frac{1}{2}(\bar{\y} - \breve{\y})^\Trans \bar{\C}_p (\bar{\y} - \breve{\y}) + \textrm{const.},
    \end{split}
\end{equation}
with $\breve{\y} = \bar{\y}' + \bar{\C}_p^{-1} \nabla p(\bar{\r}|\bar{\y}') \in \Compl^{MN}$. This can be interpreted as an additive Gaussian model given by
\begin{equation} \label{eq:1bit_real}
    \breve{\y} = \bar{\y} + \bar{\varepsilonb},
\end{equation}
with $\bar{\varepsilonb} \sim \mathcal{N}(\0, \bar{\C}_p^{-1})$. Reverting to complex numbers, we have
\begin{equation}
    \breve{\y}_\textrm{c} = \bar{\y}_\textrm{c} + \bar{\varepsilonb}_\textrm{c} \in \Compl^{MN},
\end{equation}
where the vectors are constructed by adding the first half and second half (multiplied with the imaginary unit) of the respective vectors in \eqref{eq:1bit_real}. Moreover, we have $\bar{\varepsilonb}_\textrm{c} \sim \mathcal{CN}(\0, \C_p^{-1})$, where $\C_p \in \Real^{MN \times MN}$ is the sum of the two $MN \times MN$ diagonal blocks of $\bar{\C}_p$. Substituting $\bar{\y}_\textrm{c} = g(\z)$, we have a model similar to \eqref{eq:systemmodel} but with different noise distribution. In this case, following the same logic as in Section~\ref{sec:gp_dimred}, the GP-based surrogate function is given by
\begin{equation}
    \widehat{g}_{\textrm{1-bit}}(\z) = \m_{\g} + \B (\B^\Trans \C_p \B + \C_{\widetilde{\g}\widetilde{\g}}^{-1})^{-1} \B^\Trans \C_p (\breve{\y}_\textrm{c} - \m_{\g}),
\end{equation}
which is the same as \eqref{eq:gp_pred_dimred} but with the AWGN covariance matrix $\sigma^2 \I_{MN}$ replaced with $\C_p$. The estimation proceeds as follows. We begin by initializing $\y^0 \in \Compl^{MN}$ and quadratically approximate the log-likelihood function in \eqref{eq:loglike_1bit} at this point. We then use either NL-E-SBL or NL-M-E-SBL to find an estimate $\widehat{\u}^1$, where the algorithms are modified according to the non i.i.d. distribution of the elements in the error term $\varepsilonb$. After obtaining $\widehat{\u}^1$, we set $\y^1 = \widehat{g}_{\textrm{1-bit}}(\A\widehat{\u}^1)$, and this procedure is repeated until convergence.

\section{Numerical Results} \label{sec:num}

In this section, we evaluate the performance of the proposed nonlinear estimation framework by means of simulations against different system parameters, such as the SNR, pilot length, number of antennas, number of channel paths, and strength of the LNA distortion.

\subsection{Simulation Setup} \label{sec:simulationsetup}

To measure the channel estimation accuracy, we consider the NMSE computed by averaging over 2000 independent channel realizations. As explained in Section~\ref{sec:meancov}, we estimate the channel up to a real positive scaling factor. Therefore, when computing the NMSE, we first rescale the estimate to have the same norm as the ground truth. This is justified since, in detection and beamforming, only the phases and relative amplitudes of the channel elements matter. We assume that the BS is equipped with a ULA with half-wavelength antenna spacing and generate the ground truth channels using a far-field multipath model with $L$ paths. Accordingly, the channel of UE~$k$ is given by \cite[Ch.~2.6]{Bjo17}
\begin{equation}
    \h_k = \sqrt{\frac{1}{L}}\sum_{l = 1}^L \zeta_{k,l} \a(\theta_{k,l}),
\end{equation}
where $\a(\theta_{k,l}) \in \Compl^M$ is the ULA steering vector corresponding to the angle of arrival $\theta_{k,l}$ and $\zeta_{k,l} \in \Compl$ denotes the complex path gain associated with the $l$th path. Each UE's signal propagates to the BS through $L$ distinct paths, giving rise to angular sparsity when $L$ is sufficiently small. The angles of arrival are sampled uniformly from $[\frac{\pi}{4}, \frac{3\pi}{4}]$, whereas the complex path gains are modeled as $\zeta_{k,l} \sim \setC \setN (0,1)$, $\forall k = 1, \ldots K$, $\forall l = 1, \ldots L$. In this setting, the per-antenna SNR is given by $\frac{p}{\sigma^2}$. The pilot matrix $\P$ is chosen to be orthogonal and is constructed using Zadoff-Chu (ZC) sequences \cite{frank1963polyphase,chu1972polyphase}.\footnote{In contrast to \cite{arjas2025enhanced}, which adopts DFT pilots, we use ZC pilots as they yield more stable performance in practice. This is likely due to the favorable correlation properties of ZC sequences under mild nonlinear distortions, which may lead to more reliable channel estimation and smoother convergence.} At the end, the LNA distortion in \eqref{eq:lnadistortion} is applied elementwise to the signal, followed by the addition of AWGN.

We compare the proposed nonlinear SBL methods with the LS estimator $\u_\textrm{LS} = (\A^\Herm \A)^{-1} \A^\Herm \y$, the BLMMSE estimator in \eqref{eq:BLMMSE}, and the linear SBL methods, i.e., E-SBL and M-E-SBL \cite{arjas2025enhanced}. To the best of our knowledge, there is no other suitable nonlinear estimation method that can be used as a baseline. In the BLMMSE estimator, the channel covariance matrix $\C_{\h_k} = \C_{\h}$, $\forall k = 1, \dots, K$, is computed via Monte-Carlo integration as $\C_{\h} = \mathbb{E}_{\theta}[\a(\theta) \a(\theta)^\Herm]$. The nonlinearity parameters $a_j = a$, $\forall j = 1, \dots, MN$, are assumed to be known. Given $\C_\h$ and $a$, the remaining quantities required by the BLMMSE estimator can be constructed as described in Section~\ref{sec:BLMMSE}. For the case of hybrid analog-digital beamforming, the analog combining matrix $\Q$ is constructed from the $M_\textrm{RF}$ DFT beams with the highest projection energy onto $\C_{\h}$.

Regarding the hyperparameter selection, the number of degrees of freedom is fixed to $\nu = 1$, which corresponds to the heavy-tailed Cauchy distribution. The other prior hyperparameters are set to $\gamma = \beta = 10^{-2}$, which make the estimated angular channels moderately sparse without overenforcing sparsity. The signal variance is set to $\tau^2 = \frac{10^{-2}}{\sigma^2 M}$, making it inversely proportional to the AWGN variance: this choice reduces the risk of overfitting compared with using a fixed value independent of $\sigma^2$ and $M$. For the cases of hybrid analog-digital beamforming and 1-bit ADCs, we set $\tau^2 = 10^{-2}$; for the latter, we additionally set $\gamma = \beta = 0.5$. Furthermore, we fix $\rho = 1$, which gives sufficient flexibility to model the nonlinearity while avoiding overfitting. We note that, as $\rho$ scales the distances between the inputs of the nonlinear function, its value should depend on the variance of the inputs. The GP kernel hyperparameters, namely the signal variance $\tau^2$ and the inverse length-scale $\rho^2$, are fixed empirically to ensure stable behavior across channel realizations. In principle, these parameters could also be estimated jointly with the SBL weight and scale vectors through marginal likelihood optimization or hierarchical updates: this extension is left for future work. Lastly, we use $D = 100$ pseudo-inputs (see Fig.~\ref{fig:MSE_D}), while further tuning may yield improved performance.

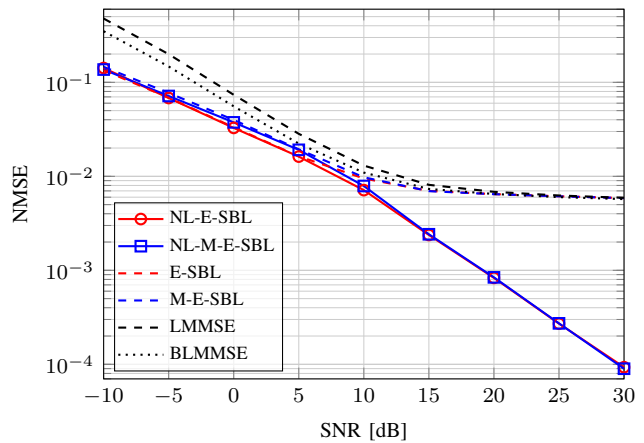
\begin{figure}[!t]
    \centering
    \begin{tikzpicture}

\begin{axis}[
	width=8.5cm,
	height=6.5cm,
	xmin=-10, xmax=30,
	ymin=0.00007, ymax=0.6,
	xlabel={SNR [dB]},
	ylabel={NMSE},
	xtick={-10,-5,0,5,10,15,20,25,30},
    extra y ticks={20},
    extra y tick labels={},
    label style={font=\footnotesize},
    ticklabel style={font=\footnotesize},
	legend style={at={(0.02,0.02)}, anchor=south west},
	legend style={font=\scriptsize, inner sep=1pt, fill opacity=0.75, draw opacity=1, text opacity=1},
	legend cell align=left,
	grid=both,
	grid style={line width=.1pt, draw=gray!40},
    ymode=log,
	title={},
	title style={font=\scriptsize, yshift=-2mm},
]

\addplot[thick, red, mark=o]
table [x=Var1, y=NMSEs_E_SBL_GP, col sep=comma] {figures_new/files_txt_new/MSE_SNR_new.txt};
\addlegendentry{NL-E-SBL};

\addplot[thick, blue, mark=square]
table [x=Var1, y=NMSEs_M_E_SBL_GP, col sep=comma] {figures_new/files_txt_new/MSE_SNR_new.txt};
\addlegendentry{NL-M-E-SBL};

\addplot[thick, red, dashed]
table [x=Var1, y=NMSEs_E_SBL, col sep=comma] {figures_new/files_txt_new/MSE_SNR_new.txt};
\addlegendentry{E-SBL};

\addplot[thick, blue, dashed]
table [x=Var1, y=NMSEs_M_E_SBL, col sep=comma] {figures_new/files_txt_new/MSE_SNR_new.txt};
\addlegendentry{M-E-SBL};

\addplot[thick, black, dashed]
table [x=Var1, y=NMSEs_LMMSE, col sep=comma] {figures_new/files_txt_new/MSE_SNR_new.txt};
\addlegendentry{LMMSE};

\addplot[thick, black, dotted]
table [x=Var1, y=NMSEs_BLMMSE, col sep=comma] {figures_new/files_txt_new/MSE_SNR_new.txt};
\addlegendentry{BLMMSE};

\end{axis}

\end{tikzpicture}
    \caption{NMSE versus SNR, with $M = 128$, $K = 5$, $N = 19$, and $\alpha = \frac{1}{3}$.}
    \label{fig:MSE_SNR}
\end{figure}

\subsection{Results} \label{sec:results}

Fig.~\ref{fig:MSE_SNR} plots the NMSE versus the SNR. At SNRs below $-5$~dB, all the SBL-based methods perform similarly, as the nonlinearity is masked by the AWGN. The performance of LS and BLMMSE is worse since they do not exploit the angular sparsity. From $10$~dB upwards, both NL-E-SBL and NL-M-E-SBL achieve considerably lower NMSE than the baselines, demonstrating the benefit of learning the distortion function. The gains are highly pronounced at $30$~dB, where the proposed nonlinear estimation framework achieves an NMSE almost two orders of magnitude lower than the baselines. These improvements occur because the GP-based surrogate function captures the nonlinear effects that dominate at high SNR, whereas the linear estimators hit an error floor. We also note that the linear SBL methods coincide with the nonlinear SBL methods at low SNR and with LS at high SNR.

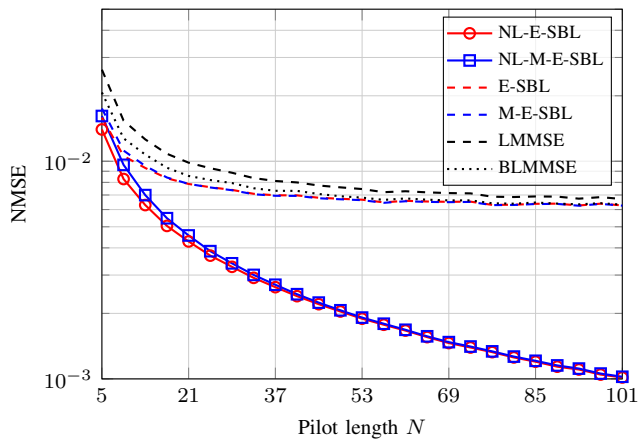
\begin{figure}[!t]
    \centering
    \begin{tikzpicture}

\begin{axis}[
	width=8.5cm,
	height=6.5cm,
	xmin=5, xmax=101,
	ymin=0.001, ymax=0.05,
	xlabel={Pilot length $N$},
	ylabel={NMSE},
	xtick={5,21,...,101},
    label style={font=\footnotesize},
    ticklabel style={font=\footnotesize},
	legend style={at={(0.98,0.98)}, anchor=north east},
	legend style={font=\scriptsize, inner sep=1pt, fill opacity=0.75, draw opacity=1, text opacity=1},
	legend cell align=left,
	grid=both,
	grid style={line width=.1pt, draw=gray!40},
    ymode=log,
	title={},
	title style={font=\scriptsize, yshift=-2mm},
]

\addplot[thick, red, mark=o]
table [x=Var1, y=NMSEs_E_SBL_GP, col sep=comma] {figures_new/files_txt_new/MSE_N_new.txt};
\addlegendentry{NL-E-SBL};

\addplot[thick, blue, mark=square]
table [x=Var1, y=NMSEs_M_E_SBL_GP, col sep=comma] {figures_new/files_txt_new/MSE_N_new.txt};
\addlegendentry{NL-M-E-SBL};

\addplot[thick, red, dashed]
table [x=Var1, y=NMSEs_E_SBL, col sep=comma] {figures_new/files_txt_new/MSE_N_new.txt};
\addlegendentry{E-SBL};

\addplot[thick, blue, dashed]
table [x=Var1, y=NMSEs_M_E_SBL, col sep=comma] {figures_new/files_txt_new/MSE_N_new.txt};
\addlegendentry{M-E-SBL};

\addplot[thick, black, dashed]
table [x=Var1, y=NMSEs_LMMSE, col sep=comma] {figures_new/files_txt_new/MSE_N_new.txt};
\addlegendentry{LMMSE};

\addplot[thick, black, dotted]
table [x=Var1, y=NMSEs_BLMMSE, col sep=comma] {figures_new/files_txt_new/MSE_N_new.txt};
\addlegendentry{BLMMSE};

\end{axis}

\end{tikzpicture}
    \caption{NMSE versus pilot length, with $\textrm{SNR} = 12$~dB, $M = 128$, $K = 5$, and $\alpha = \frac{1}{3}$.}
    \label{fig:MSE_N}
\end{figure}

Fig.~\ref{fig:MSE_N} shows the NMSE versus the pilot length $N$. As the pilot length grows, the NMSE is reduced for all the estimators, but with distinct slopes. The proposed nonlinear SBL methods are superior for all pilot lengths, with performance gap increasing with $N$. While longer pilot sequences benefit all the estimators, the proposed nonlinear SBL methods achieve the largest NMSE improvements and delay the onset of the error floor more effectively than the baseline methods. These trends mirror those observed for varying SNR, since increasing the pilot length and boosting the SNR reduce the estimation error in similar ways.

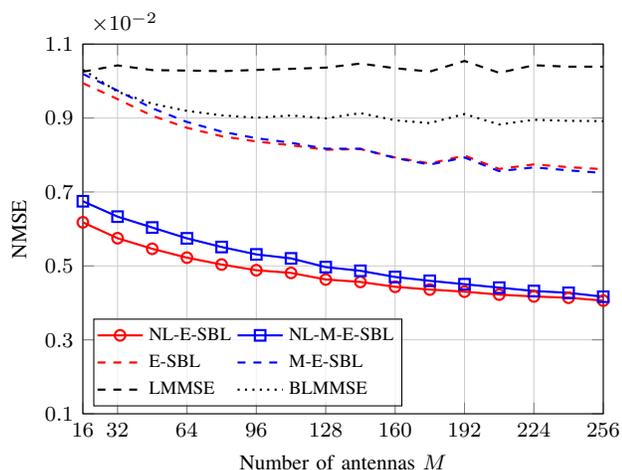
\begin{figure}[!t]
    \centering
    \begin{tikzpicture}

\begin{axis}[
	width=8.5cm,
	height=6.5cm,
	xmin=16, xmax=256,
	ymin=0.001, ymax=0.011,    
	xlabel={Number of antennas $M$},
	ylabel={NMSE},
	xtick={16,32,64,...,256},
    ytick={0.001,0.003,...,0.011},
    tick scale binop=\times,
    label style={font=\footnotesize},
    ticklabel style={font=\footnotesize},
    legend columns=2,
	legend style={at={(0.02,0.02)}, anchor=south west},
	legend style={font=\scriptsize, inner sep=1pt, fill opacity=0.75, draw opacity=1, text opacity=1},
	legend cell align=left,
	grid=both,
	grid style={line width=.1pt, draw=gray!40},
	title={},
	title style={font=\scriptsize, yshift=-2mm}
]

\addplot[thick, red, mark=o]
table [x=Var1, y=NMSEs_E_SBL_GP, col sep=comma] {figures_new/files_txt_new/MSE_M_new.txt};
\addlegendentry{NL-E-SBL};

\addplot[thick, blue, mark=square]
table [x=Var1, y=NMSEs_M_E_SBL_GP, col sep=comma] {figures_new/files_txt_new/MSE_M_new.txt};
\addlegendentry{NL-M-E-SBL};

\addplot[thick, red, dashed]
table [x=Var1, y=NMSEs_E_SBL, col sep=comma] {figures_new/files_txt_new/MSE_M_new.txt};
\addlegendentry{E-SBL};

\addplot[thick, blue, dashed]
table [x=Var1, y=NMSEs_M_E_SBL, col sep=comma] {figures_new/files_txt_new/MSE_M_new.txt};
\addlegendentry{M-E-SBL};

\addplot[thick, black, dashed]
table [x=Var1, y=NMSEs_LMMSE, col sep=comma] {figures_new/files_txt_new/MSE_M_new.txt};
\addlegendentry{LMMSE};

\addplot[thick, black, dotted]
table [x=Var1, y=NMSEs_BLMMSE, col sep=comma] {figures_new/files_txt_new/MSE_M_new.txt};
\addlegendentry{BLMMSE};

\end{axis}

\end{tikzpicture}
    \caption{NMSE versus number of antennas, with $\textrm{SNR} = 12$~dB, $K = 5$, $N = 19$, and $\alpha = \frac{1}{3}$.}
    \label{fig:MSE_M}
\end{figure}

Fig.~\ref{fig:MSE_M} illustrates the NMSE versus the number of antennas $M$. As the number of antennas increases, all the estimators except LS exhibit steadily decreasing NMSE, but with different slopes and error floors. The proposed nonlinear SBL methods lead throughout, with NL-E-SBL performing slightly better than NL-M-E-SBL at small antenna counts (though this gap narrows as $M$ grows). These results confirm that, while more antennas improve the performance of all the estimators, embedding the enhanced SBL methods into the GP-based framework consistently delivers the highest estimation accuracy. In addition, more significant gains can be obtained for higher SNR and longer pilots (see Fig.~\ref{fig:MSE_SNR} and Fig.~\ref{fig:MSE_N}).

\begin{figure}[!t]
    \centering
    \begin{tikzpicture}

\begin{axis}[
	width=8.5cm,
	height=6.5cm,
	xmin=1, xmax=10,
	ymin=0.002, ymax=0.012,
	xlabel={Number of channel paths $L$},
	ylabel={NMSE},
	xtick={1,2,3,4,5,6,7,8,9,10},
    ytick={0.002,0.004,...,0.012},
    tick scale binop=\times,
    label style={font=\footnotesize},
    ticklabel style={font=\footnotesize},
    legend columns=2,
	legend style={at={(0.98,0.02)}, anchor=south east},
	legend style={font=\scriptsize, inner sep=1pt, fill opacity=0.75, draw opacity=1, text opacity=1},
	legend cell align=left,
	grid=both,
	grid style={line width=.1pt, draw=gray!40},
	title={},
	title style={font=\scriptsize, yshift=-2mm},
]

\addplot[thick, red, mark=o]
table [x=Var1, y=NMSEs_E_SBL_GP, col sep=comma] {figures_new/files_txt_new/MSE_L_new.txt};
\addlegendentry{NL-E-SBL};

\addplot[thick, blue, mark=square]
table [x=Var1, y=NMSEs_M_E_SBL_GP, col sep=comma] {figures_new/files_txt_new/MSE_L_new.txt};
\addlegendentry{NL-M-E-SBL};

\addplot[thick, red, dashed]
table [x=Var1, y=NMSEs_E_SBL, col sep=comma] {figures_new/files_txt_new/MSE_L_new.txt};
\addlegendentry{E-SBL};

\addplot[thick, blue, dashed]
table [x=Var1, y=NMSEs_M_E_SBL, col sep=comma] {figures_new/files_txt_new/MSE_L_new.txt};
\addlegendentry{M-E-SBL};

\addplot[thick, black, dashed]
table [x=Var1, y=NMSEs_LMMSE, col sep=comma] {figures_new/files_txt_new/MSE_L_new.txt};
\addlegendentry{LMMSE};

\addplot[thick, black, dotted]
table [x=Var1, y=NMSEs_BLMMSE, col sep=comma] {figures_new/files_txt_new/MSE_L_new.txt};
\addlegendentry{BLMMSE};

\end{axis}

\end{tikzpicture}
    \caption{NMSE versus number of propagation paths, with $\textrm{SNR} = 12$~dB, $M = 128$, $K = 5$, $N = 19$, and $\alpha = \frac{1}{3}$.}
    \label{fig:MSE_L}
\end{figure}
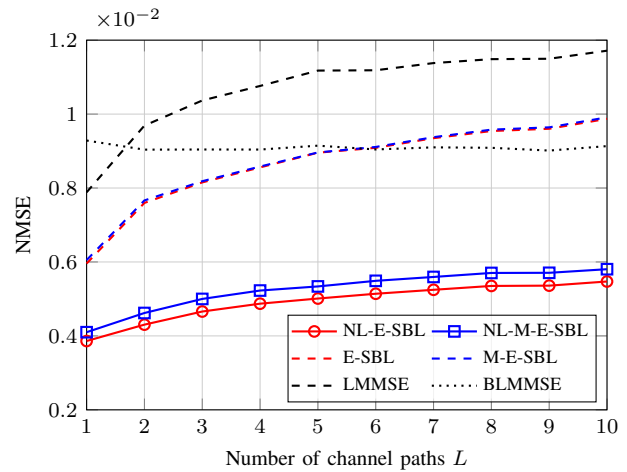

Fig.~\ref{fig:MSE_L} plots the NMSE versus the number of channel paths $L$. All the estimators except BLMMSE suffer increasing error as the scattering complexity grows. The proposed nonlinear SBL methods offer the best estimation accuracy throughout. For $L = 1$, the NMSE with BLMMSE is around $140\%$ higher than with NL-E-SBL, decreasing to $67\%$ at $L = 10$. This highlights that, while BLMMSE remains unaffected by the number of paths due to its fixed covariance assumption, it is consistently outperformed by the proposed nonlinear SBL methods, which adapt more effectively to the underlying channel structure.

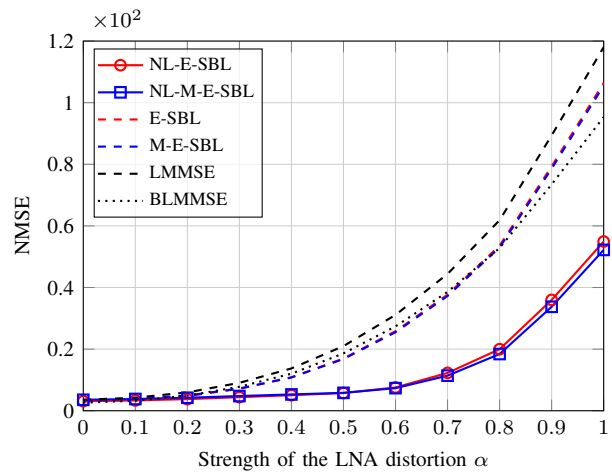
\begin{figure}[!t]
    \centering
    \begin{tikzpicture}

\begin{axis}[
	width=8.5cm,
	height=6.5cm,
	xmin=0, xmax=1,
	ymin=0, ymax=0.12,
	xlabel={Strength of the LNA distortion $\alpha$},
	ylabel={NMSE},
	xtick={0,0.1,0.2,0.3,0.4,0.5,0.6,0.7,0.8,0.9,1},
	ytick={0,0.02,...,0.12},
    scaled y ticks = base 10:1,
    tick scale binop=\times,
	yticklabels={0,0.2,0.4,0.6,0.8,1,1.2},
    label style={font=\footnotesize},
    ticklabel style={font=\footnotesize},
	legend style={at={(0.02,0.98)}, anchor=north west},
	legend style={font=\scriptsize, inner sep=1pt, fill opacity=0.75, draw opacity=1, text opacity=1},
	legend cell align=left,
	grid=both,
	grid style={line width=.1pt, draw=gray!40},
	title={},
	title style={font=\scriptsize, yshift=-2mm},
]

\addplot[thick, red, mark=o]
table [x=Var1, y=NMSEs_E_SBL_GP, col sep=comma] {figures_new/files_txt_new/MSE_alpha_new.txt};
\addlegendentry{NL-E-SBL};

\addplot[thick, blue, mark=square]
table [x=Var1, y=NMSEs_M_E_SBL_GP, col sep=comma] {figures_new/files_txt_new/MSE_alpha_new.txt};
\addlegendentry{NL-M-E-SBL};

\addplot[thick, red, dashed]
table [x=Var1, y=NMSEs_E_SBL, col sep=comma] {figures_new/files_txt_new/MSE_alpha_new.txt};
\addlegendentry{E-SBL};

\addplot[thick, blue, dashed]
table [x=Var1, y=NMSEs_M_E_SBL, col sep=comma] {figures_new/files_txt_new/MSE_alpha_new.txt};
\addlegendentry{M-E-SBL};

\addplot[thick, black, dashed]
table [x=Var1, y=NMSEs_LMMSE, col sep=comma] {figures_new/files_txt_new/MSE_alpha_new.txt};
\addlegendentry{LMMSE};

\addplot[thick, black, dotted]
table [x=Var1, y=NMSEs_BLMMSE, col sep=comma] {figures_new/files_txt_new/MSE_alpha_new.txt};
\addlegendentry{BLMMSE};

\end{axis}

\end{tikzpicture}
    \caption{NMSE versus strength of the LNA distortion, with $\textrm{SNR} = 12$~dB, $M = 128$, $K = 5$, and $N = 19$.}
    \label{fig:MSE_a}
\end{figure}

Fig.~\ref{fig:MSE_a} illustrates the impact of the strength of the LNA distortion $\alpha$ on the NMSE. As the LNAs become less linear, all the methods incur higher error, though with different sensitivities. The proposed nonlinear SBL methods remain the most robust: their NMSE increases from $0.003$ at $\alpha = 0$ (no impairments) to around $0.05$ at $\alpha = 1$ (severe impairments). This behavior is expected: as the nonlinearities grow stronger, the underlying optimization problem becomes more nonconvex and increasingly sensitive to initialization, which reduces robustness. Despite this, the proposed algorithms still achieve substantial performance gains even for strong nonlinearities. At $\alpha = 1$, BLMMSE slightly outperforms the other linear methods, but the differences are otherwise negligible.

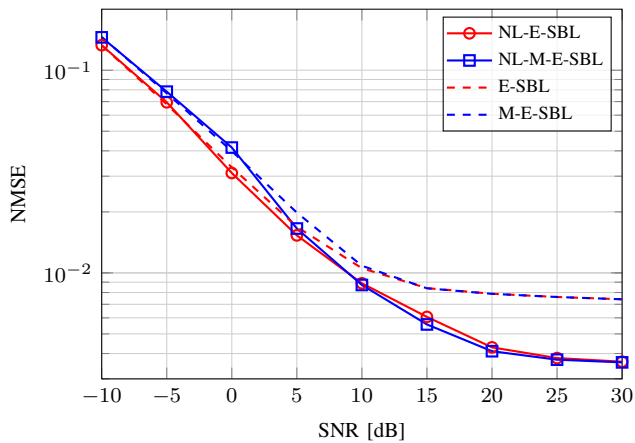
\begin{figure}[!t]
    \centering
    \begin{tikzpicture}

\begin{axis}[
	width=8.5cm,
	height=6.5cm,
	xmin=-10, xmax=30,
	ymin=0.003, ymax=0.2,
	xlabel={SNR [dB]},
	ylabel={NMSE},
	xtick={-10,-5,0,5,10,15,20,25,30},
    label style={font=\footnotesize},
    ticklabel style={font=\footnotesize},
	legend style={at={(0.98,0.98)}, anchor=north east},
	legend style={font=\scriptsize, inner sep=1pt, fill opacity=0.75, draw opacity=1, text opacity=1},
	legend cell align=left,
	grid=both,
	grid style={line width=.1pt, draw=gray!40},
    ymode=log,
	title={},
	title style={font=\scriptsize, yshift=-2mm},
]

\addplot[thick, red, mark=o]
table [x=Var1, y=NMSEs_E_SBL_GP, col sep=comma] {figures_new/files_txt_new/MSE_SNR_HB_new_1.txt};
\addlegendentry{NL-E-SBL};

\addplot[thick, blue, mark=square]
table [x=Var1, y=NMSEs_M_E_SBL_GP, col sep=comma] {figures_new/files_txt_new/MSE_SNR_HB_new_1.txt};
\addlegendentry{NL-M-E-SBL};

\addplot[thick, red, dashed]
table [x=Var1, y=NMSEs_E_SBL, col sep=comma] {figures_new/files_txt_new/MSE_SNR_HB_new_1.txt};
\addlegendentry{E-SBL};

\addplot[thick, blue, dashed]
table [x=Var1, y=NMSEs_M_E_SBL, col sep=comma] {figures_new/files_txt_new/MSE_SNR_HB_new_1.txt};
\addlegendentry{M-E-SBL};

\end{axis}

\end{tikzpicture}
    \caption{NMSE versus SNR with hybrid analog-digital beamforming, with $M = 128$, $M_\textrm{RF} = 96$, $K = 5$, and $N = 19$.}
    \label{fig:MSE_hb}
\end{figure}

Fig.~\ref{fig:MSE_hb} shows the NMSE versus the SNR with hybrid analog-digital beamforming (see Section~\ref{sec:special_cases_hb}). The trend differs from the fully digital case shown in Fig.~\ref{fig:MSE_SNR}, as the proposed nonlinear SBL methods exhibit an error floor around $25$~dB. Nevertheless, they achieve lower error than their linear counterparts in the high SNR regime. At SNRs below $5$~dB, all the methods perform similarly due to the strong AWGN masking the nonlinearity.

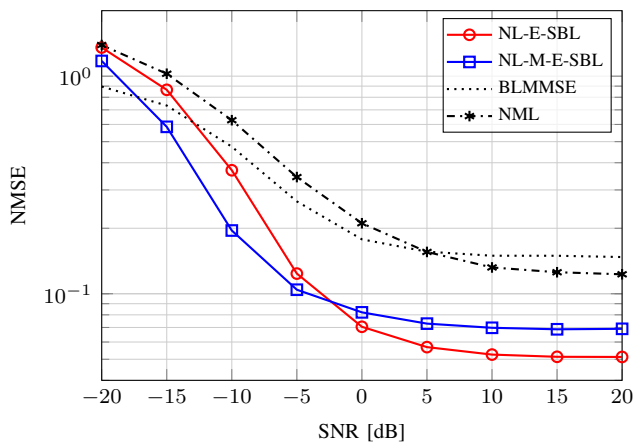
\begin{figure}[!t]
    \centering
    \begin{tikzpicture}

\begin{axis}[
	width=8.5cm,
	height=6.5cm,
	xmin=-20, xmax=20,
	ymin=0.04, ymax=2,
	xlabel={SNR [dB]},
	ylabel={NMSE},
	xtick={-20,-15,-10,-5,0,5,10,15,20},
    ytick={-10,0},
    label style={font=\footnotesize},
    ticklabel style={font=\footnotesize},
	legend style={at={(0.98,0.98)}, anchor=north east},
	legend style={font=\scriptsize, inner sep=1pt, fill opacity=0.75, draw opacity=1, text opacity=1},
	legend cell align=left,
	grid=both,
	grid style={line width=.1pt, draw=gray!40},
    ymode=log,
	title={},
	title style={font=\scriptsize, yshift=-2mm},
]

\addplot[thick, red, mark=o]
table [x=Var1, y=NMSEs_E_SBL_GP, col sep=comma] {figures_new/files_txt_new/MSE_SNR_1bit_new_1.txt};
\addlegendentry{NL-E-SBL};

\addplot[thick, blue, mark=square]
table [x=Var1, y=NMSEs_M_E_SBL_GP, col sep=comma] {figures_new/files_txt_new/MSE_SNR_1bit_new_1.txt};
\addlegendentry{NL-M-E-SBL};

\addplot[thick, black, dotted]
table [x=Var1, y=NMSEs_BLMMSE, col sep=comma] {figures_new/files_txt_new/MSE_SNR_1bit_new_1.txt};
\addlegendentry{BLMMSE};

\addplot[thick, black, dash dot,mark=asterisk, mark options={solid}]
table [x=Var1, y=NMSEs_NML, col sep=comma] {figures_new/files_txt_new/MSE_SNR_1bit_new_1.txt};
\addlegendentry{NML};

\end{axis}

\end{tikzpicture}
    \caption{NMSE versus SNR with 1-bit ADCs, with $M = 128$, $K = 5$, and $N = 19$.}
    \label{fig:MSE_1bit}
\end{figure}

Lastly, Fig.~\ref{fig:MSE_1bit} plots the NMSE versus the SNR with 1-bit ADCs (see Section~\ref{sec:special_cases_1bit}), comparing the proposed nonlinear estimation framework with the BLMMSE estimator for 1-bit ADCs from \cite{li2017channel} and the near-ML (NML) estimator from \cite{choi2016near}. The proposed NL-E-SBL and NL-M-E-SBL consistently outperform both BLMMSE and NML across most of the SNR range. At low SNR, NL-M-E-SBL outperforms NL-E-SBL, while the ranking is reversed at around $-2$~dB. Although hybrid analog-digital beamforming and 1‑bit ADCs complicate the estimation of the LNA distortion, these results show the effectiveness of the proposed nonlinear estimation framework.

As a general comment, the gains of BLMMSE over the other linear estimators are largely diminished in our setting. Because the nonlinearity acts independently at each antenna, the Bussgang gain reduces to a diagonal scaling factor. As discussed in Section~\ref{sec:simulationsetup}, all the estimators are subsequently rescaled to match the norm of the ground truth, which removes this scaling advantage. At high SNR, the remaining error is dominated by modeling mismatch rather than amplitude bias, which explains the nearly overlapping performance of BLMMSE and the other linear estimators observed in Fig.~\ref{fig:MSE_SNR}. Moreover, since BLMMSE does not leverage sparsity, E-SBL and M-E-SBL from \cite{arjas2025enhanced} (which rely on compressive sensing) retain a clear performance advantage at low SNR.

\begin{figure*}
\begin{align} \label{eq:proof}
\begin{split}
\bar f_\u(\widetilde{\u}^{t+1}) & \le \bar f_\u(\widetilde{\u}^t) + \mathrm{Re}\left[\nabla \bar f_\u(\widetilde{\u}^t)^\Herm(\widetilde{\u}^{t+1}-\widetilde{\u}^t)\right]+ \frac{L^t}{2}\|\widetilde{\u}^{t+1}-\widetilde{\u}^t\|^2 \\
& =\bar f_\u(\widetilde{\u}^t) - \delta\,\mathrm{Re}\big[  \bigl(\nabla \bar f_\u^t(\widehat{\u}^t)-\nabla \bar f_\u^t(\widetilde{\u}^t)\bigr)^\Herm (\widehat{\u}^t-\widetilde{\u}^t) \big] + \frac{L^t\delta^2}{2}\|\widehat{\u}^t -\widetilde{\u}^t\|^2
\end{split}
\end{align}
\hrulefill
\vspace{-4mm}
\end{figure*}

\subsection{Discussion} \label{sec:discussion}

In all scenarios, the proposed methods outperform BLMMSE, which has traditionally been used for channel estimation in the presence of hardware impairments. This improvement stems from the fact that the proposed nonlinear estimation framework explicitly models the distortion function, while BLMMSE only captures its first- and second-order statistics. A further advantage of the framework is that it does not require prior knowledge of the mathematical form of the distortion function or its statistics. On the other hand, BLMMSE has the benefits of being non-iterative and more amenable to theoretical performance analysis.

The potential use of the proposed nonlinear SBL methods is not limited to channel estimation, and they can be considered as a part of a broader context of GP-based statistical learning. More specifically, recent research on GPs has largely focused on scalability and dimensionality reduction (see e.g., \cite{tripathy2016gaussian,snelson2012variable,liu2017dimension,binois2022survey,gramacy2012gaussian,eriksson2021high}). The proposed nonlinear estimation framework can be seen as a novel contribution to this literature, as it combines the use of pseudo-inputs to improve scalability and sparsity to reduce the effective dimension. Conventionally, parameter estimation using GPs is done by maximum marginal likelihood, whereas we minimize the penalized $\ell_2$-norm between the GP prediction and the observed data. To maximize the marginal likelihood, the GN method cannot be directly applied, and quasi-Newton methods such as the Broyden-Fletcher-Goldfarb-Shanno (BFGS) algorithm are usually utilized. However, for mildly nonlinear problems and given a good initialization, BFGS typically converges more slowly than GN \cite{nocedal1999numerical}. Hence, in time-sensitive applications, our approach of minimizing the $\ell_2$-norm is generally more efficient.

One limitation of using the $\ell_2$-norm for parameter estimation is that it relies solely on the GP's posterior mean, thereby ignoring the predictive uncertainty captured by the GP's posterior variance. This uncertainty reflects the confidence of the model in its predictions, particularly in regions with limited data or strong noise. However, in the context of linear data detection, the symbols are detected based on a given channel estimate, for which the posterior mean alone often suffices (as uncertainty information does not directly influence the detection rule). Moreover, although objectives based on $\ell_2$-norm can lead to overfitting, especially in high-dimensional settings or with limited data, we address this risk by introducing a sparsifying prior. This promotes solutions with fewer active components, effectively acting as a regularizer that limits the model's complexity and improves generalization.

Lastly, the joint estimation of the channel and the nonlinear function raises an identifiability question. In principle, identifiability requires $g$ to be injective, i.e., $g(\mathbf{z}_1) \neq g(\mathbf{z}_2)$ whenever $\mathbf{z}_1 \neq \mathbf{z}_2$. While establishing injectivity is generally intractable for flexible, data-driven models like GPs, we can mitigate potential identifiability issues by constraining the nonlinearity through the GP hyperparameters. Specifically, one can model $g$ as $g(z) = z + \tau^2 \varepsilon(\rho^2 z)$, where $\varepsilon$ is a nonlinear function: as $\tau^2 \to 0$, the function approaches an identity mapping, thereby promoting injectivity; similarly, smaller values of $\rho^2$ stiffen $g$, effectively reducing its Lipschitz constant and limiting local variations that could otherwise impair injectivity. This structured modeling choice improves the practical identifiability of the latent input, even without formal guarantees. Despite the absence of a provable guarantee, the numerical results demonstrate robust performance over a wide range of system parameters.

\section{Conclusions} \label{sec:concl}

We proposed a nonlinear channel estimation framework that models the distortion function arising from hardware impairments using GP regression while leveraging the inherent sparsity of massive MIMO channels. The resulting nonlinear SBL methods achieve significant NMSE reduction compared with LS, BLMMSE, and linear SBL, particularly under strong LNA distortion and at high SNR. This advancement enables more robust beamforming in hardware‑impaired massive MIMO systems. Future work will investigate reducing the cubic complexity in the number of antennas, potentially via low‑rank Jacobian approximations or deep unfolding approaches that learn efficient optimizer iterations.

\appendix

\begin{proposition}
The NL-M-E-SBL updates yield a monotonically increasing value of the objective in \eqref{eq:obj_func} provided that the step size satisfies 
\[
0 < \delta < \frac{2\mu^t}{L^t},
\]
with $\mu^t$ and $L^t$ defined below.
\end{proposition}

\begin{IEEEproof}
\begin{itemize}
    \item[$\bullet$] For consistency with convex optimization conventions, we analyze the negative objective $\bar f(\widetilde \u, \w, \s) = -f(\widetilde \u, \w, \s)$ (see \eqref{eq:obj_func}).
    \item[$\bullet$] With the squared exponential kernel, the map $\widehat g_{\mathrm r}$ is Lipschitz smooth with constant $L^t>0$.
    \item[$\bullet$] As $\widetilde{\W}^t$ is a diagonal matrix with positive elements, the surrogate
    \[
    \bar f_\u^t(\widetilde{\u})
    = \frac{1}{\sigma^2}\big\|\widetilde{\y}
    - \widetilde{\widehat g_\textrm{r}(\A\u^t)}
    - \widetilde{\J}_{\z^t}(\widetilde{\u}-\widetilde{\u}^t)\big\|^2
    + \widetilde{\u}^\Herm \widetilde{\W}^t \widetilde{\u}
    \]
    is strongly convex in $\widetilde{\u}$ with constant $\mu^t>0$.
\end{itemize}

\smallskip

\textit{Step 1: Descent for the $\u$-update.} Fix $\w^t$ and $\s^t$, and recall the $\u$-dependent part of the objective
\[
\bar f_\u(\widetilde{\u})
=\frac{1}{\sigma^2}\|\widetilde{\y}-\widetilde{\widehat g_\textrm{r}(\A\u)}\|^2
+ \widetilde{\u}^\Herm\widetilde{\W}^t \widetilde{\u}.
\]
The Wirtinger gradients  satisfy
\[
\nabla \bar f_\u(\widetilde{\u}^t)=\nabla \bar f_\u^t(\widetilde{\u}^t),
\qquad 
\nabla \bar f_\u^t(\widehat{\u}^t)=\0,
\]
with $\widehat{\u}^t=\frac{1}{\sigma^2} \Sigmab^t \widetilde{\J}_{\z^t}^\Herm(\widetilde{\y} - \m_{\widetilde{\y}}^t)$. The damped update is
\[
\widetilde{\u}^{t+1}
=\delta\widehat{\u}^t + (1-\delta)\widetilde{\u}^t.
\]
Because $\nabla \bar f_\u$ is $L^t$-Lipschitz, the descent lemma gives \eqref{eq:proof} at the top of the page. The strong convexity of $\bar f_\u^t$ implies
\[
\mathrm{Re}\big[\bigl(\nabla \bar f_\u^t(\widehat{\u}^t)-\nabla \bar f_\u^t(\widetilde{\u}^t)\bigr)^\Herm (\widehat{\u}^t-\widetilde{\u}^t)\big] \ge \mu^t\|\widehat{\u}^t-\widetilde{\u}^t\|^2.
\]
Therefore, we have
\[
\bar f_\u(\widetilde{\u}^{t+1}) \le \bar f_\u(\widetilde{\u}^t) - \delta\left(\mu^t-\frac{L^t\delta}{2}\right)\|\widehat{\u}^t-\widetilde{\u}^t\|^2,
\]
where the last term is positive for $0 < \delta < 2\mu^t/L^t$.

\smallskip

\textit{Step 2: Descent for the $\w$-update.} Fix $\widetilde{\u}^{t+1}$ and $\s^t$, and define $\omega_j=\log w_j$, $\forall j = 1, \dots, MK$, and $\omegab =~[\omega_1 \dots \omega_{MK}]^\Trans \in \Real^{MK}$. Recall the $\omegab$-dependent part of the objective
{\small
\[
\bar f_\w(\omegab) = \sum_{j = 1}^{MK} \bigg( 2\log(e^{\omega_j} s_j) + 2\frac{|u_j|^2}{e^{\omega_j} s_j} + \frac{\nu + 2}{2}\omega_j + \frac{\nu}{2e^{\omega_j}} \bigg),
\]
}
Its second derivative is
\[
\frac{\partial^2 \bar f_\w}{\partial\omega_j^2} = 2\frac{|u_j^{t+1}|^2}{e^{\omega_j}s_j} + \frac{\nu}{2e^{\omega_j}} > 0, \ \forall j = 1, \dots, MK,
\]
establishing strict convexity. Thus, $\bar f_\w$ has a unique minimizer, and the closed-form update
\[
e^{\omega_j^{t+1}} = w_j^{t+1} = \frac{2|u_j^{t+1}|^2/s_j + \nu/2}{\nu/2 + 3}
\]
satisfies $\bar f_\w(\omegab^{t+1}) < \bar f_\w(\omegab^t)$.

\smallskip

\textit{Step 3: Descent for the $\s$-update.} Each $s_j$ has the same inverse gamma prior family as $w_j$, and the corresponding log-domain objective is strictly convex. Hence, its minimizer also strictly decreases the objective.

\smallskip

Combining steps 1 to 3 yields monotone ascent for the objective \eqref{eq:obj_func}
\[
f(\u^{t+1},\w^{t+1},\s^{t+1}) > f(\u^t,\w^t,\s^t),
\]
which establishes the claim.
\end{IEEEproof}

\bibliographystyle{IEEEtran}
\bibliography{refs_abbr,bibliography}

\end{document}